\documentclass[twocolumn]{article}
\usepackage[utf8]{inputenc}
\usepackage[margin=16mm]{geometry}
\usepackage[skip=3pt]{parskip}
\usepackage[dvipsnames]{xcolor}
\usepackage[colorlinks=true, allcolors=Maroon]{hyperref}

\usepackage{kpfonts}
\usepackage{physics}
\usepackage{cleveref}
\usepackage{tabularx} 
\usepackage{adjustbox}

\usepackage{authblk}

\usepackage{datetime2}
\usepackage{amsmath}
\usepackage{algorithm}
\usepackage{algorithmic}

\usepackage{orcidlink}
\usepackage{wasysym}
\usepackage{float}

\usepackage{tabulary}
\usepackage{booktabs, amsfonts}
\usepackage{multirow}

\usepackage{graphicx}

\usepackage{babel}

\usepackage{subcaption}

\newcolumntype{M}[1]{>{\centering\arraybackslash}m{#1}}

\title{A Resource Allocating Compiler for Lattice Surgery}
\author[1]{Alan Robertson \thanks{alan.robertson@uts.edu.au}}
\author[1]{Haowen Gao}
\author[1]{Yuval R.~Sanders \orcidlink{0000-0001-8003-0039} \thanks{yuval.sanders@uts.edu.au}}
\affil[1]{%
  Centre for Quantum Software and Information,
  University of Technology Sydney,
  NSW 2007, Australia}
\date{\footnotesize Compilation timestamp: \DTMnow}

\begin{document}

\maketitle

\begin{abstract}
  The emerging field of quantum resource estimation is aimed at providing estimates
  of the hardware requirements (`quantum resources') needed to execute a useful,
  fault-tolerant quantum computation. 
  Given that quantum computers are intended to compete
  with supercomputers, useful quantum computations are likely to involve the use of millions
  of qubits and error correction clock cycles.
  The compilation and benchmarking of these circuits depends on placement and routing algorithms, 
  which are infeasible to construct at scale by hand. 
  We offer a compiler that transforms a quantum circuit into a sequence of lattice surgery operations.
  The compiler manages memory in terms of surface code patches and costs the space-time volume and cycle counts of the input circuits. 
  These compiled lattice surgery objects are then recursively repurposed as gates for larger scale compilations.    
  Our code is available on GitHub under a permissive software license and we welcome community contributions.
\end{abstract}


\section{Introduction}
\label{sec:introduction}

In recent decades, quantum computers have developed from a purely theoretical
computational model to one that could be implemented at a large scale with near-future technology. 
This state of affairs has come about due to developments
in both theoretical advances in fault-tolerant quantum computing architectures
as well as technological advances in quantum information processing hardware.
Even more recently, the need has arisen to perform `quantum resource estimation',
which is the process of estimating the hardware requirements for a large quantum computation.

Previous developments in this direction have typically been architecture agnostic~\cite{AQREpaper}, or have required pre-determined surface code layouts~\cite{opensurgery,Leblond_2024}.      
In the agnostic case costs are determined as a factor of the most expensive operation in the underlying gate-set, while hand-compiled surface code layouts admit the incorporation of architecturally relevant costs. 
For example, without a specific layout of register patches, routing, route contention and idling costs are neglected.

In the case of hand-compiled layouts, the costs are sensitive to the placement choices made by the programmer, which requires      
In this paper, we describe a lattice surgery compiler that incorporates a number of heuristics that tackle automatic qubit placement, routing and more generic resource allocation patterns. 
These placement patterns form a recursive allocation structure, allowing pre-compiled program elements to act as both gates within the DAG and closures over regions of memory on the surface code.

We compile and discuss the performance of this strategy over a range of common quantum primitives and algorithms. 

\subsection{Contributions}
\label{sec:introduction/contributions}

In this manuscript, we present code
(\href{https://github.com/Alan-Robertson/Surface\_Code\_Compiler}{available on GitHub})
that accepts a description of a quantum circuit and a device specification
but returns an explicit sequence of lattice surgery instructions for the given device.

Our compiler does not make approximations or estimates of the space-time and run-time costs of the execution of a given algorithm.  
Instead these costs are supported by a full implementation, including memory constraints imposed by the hardware.   
This contrasts with costing methodologies that rely on simplifying assumptions such as arbitrary connectivity, infinite memory for $T$ factories and distilleries, ignored idling costs, or excluded ancillae dependency contentions.    

At this stage we have only implemented one device model;
namely, the `game of surface codes' lattice surgery ISA of Litinski~\cite{gameofsurfacecodes},
but we intend to expand the functionality of the code to incorporate other ISAs.
The design of the code is therefore similar to other compilers in that the compilation
process consists of four broad stages described below.
\begin{enumerate}
  \item \textbf{Parsing.}
  In our case, the parsing step translates the input into a partially ordered set of
  quantum operations that is represented as a directed acyclic graph (DAG).
  The nodes of the DAG are of three kinds:
  native operations, 
  macro-instructions (`macros'),
  and external elements (`externs').

  \item \textbf{Allocation.}
  Here the job is to assign specific hardware locations (`registers') for storage
  and manipulation of logical qubits. The allocation task must also incorporate
  reserved space for externs as well as unallocated space (`buses') to be used during
  program execution to enable joint operations between distant registers and/or externs.
  
  \item \textbf{Mapping.}
  Having allocated hardware locations to be used as registers, buses, or for externs,
  the next task is to assign specific logical qubits to specific registers.
  We design a heuristic that is intended to keep the overall circuit execution time
  as small as possible, though we note that such a problem is in general NP-hard.
  
  \item \textbf{Routing.}
  While mapping allocates fixed resources to locations on the hardware mock-up, it does not dictate how ancillae should be allocated to computational operations. The routing pass handles ancillae for multi-qubit operations, and lattice surgery operations that require additional patches.      
\end{enumerate}

We evaluate the implementation of our compiler against a variety of circuit inputs.   
(1) basic quantum logic such as CNOT networks and magic state factories,
(2) common primitive elements in quantum algorithms such as
    multi-controlled NOT operations and the quantum Fourier transform,
(3) coherent arithmetic operations (i.e.~integer addition and multiplication), and
(4) quantum memory -- specifically, bucket-brigade QRAM, QROM.

\subsection{Terminology}
\label{sec:introduction/terminology}

We introduce several technical concepts in this paper to aid in the
design and implementation of our compiler. At a high level, we view the
compilation challenge as a `qubit mapping' problem, in which we must
choose for each logical qubit some memory location in a large
surface code quantum computer. This allocation of qubits should seek to
minimise space-time volume overheads due to routing.

Routing overheads are incurred when gates contend for ancillae resources, such as routing ancillae for multi-qubit gates, or movement operations for $T$ states.    
These operations require contiguous regions of ancillae states, which may introduce implicit dependencies between otherwise independent operations.     
Space-time volume overheads incurred by routing include both the route distance, and cycles lost to idling qubits waiting for free ancillae resources.

We then model the compilation in terms of the specification of
four kinds of regions (`patches') in quantum memory: register patches,
routing patches, extern patches and IO patches. 
Register patches host logical qubits, and are mapped by name.
Routing patches act as a bus and provide ancillae support for both local and non-local operations.
Extern patches and IO patches act to define and join logical computational units.

These patches taken together comprise a territory we call a
\emph{quantum circuit board}~(QCB), because the laying out of these
patches is conceptually similar to the process of laying out of components on a classical circuit board.

The output of our compilation algorithm is a QCB: a data structure specifying a set of surface code
patches, an allocation of logical qubits to some of those patches,
and a sequence of the lattice surgery operations~\cite{gameofsurfacecodes}
needed to execute the input quantum circuit.

\begin{figure*}
\center
\begin{subfigure}{\textwidth}
\includegraphics[scale=0.65]{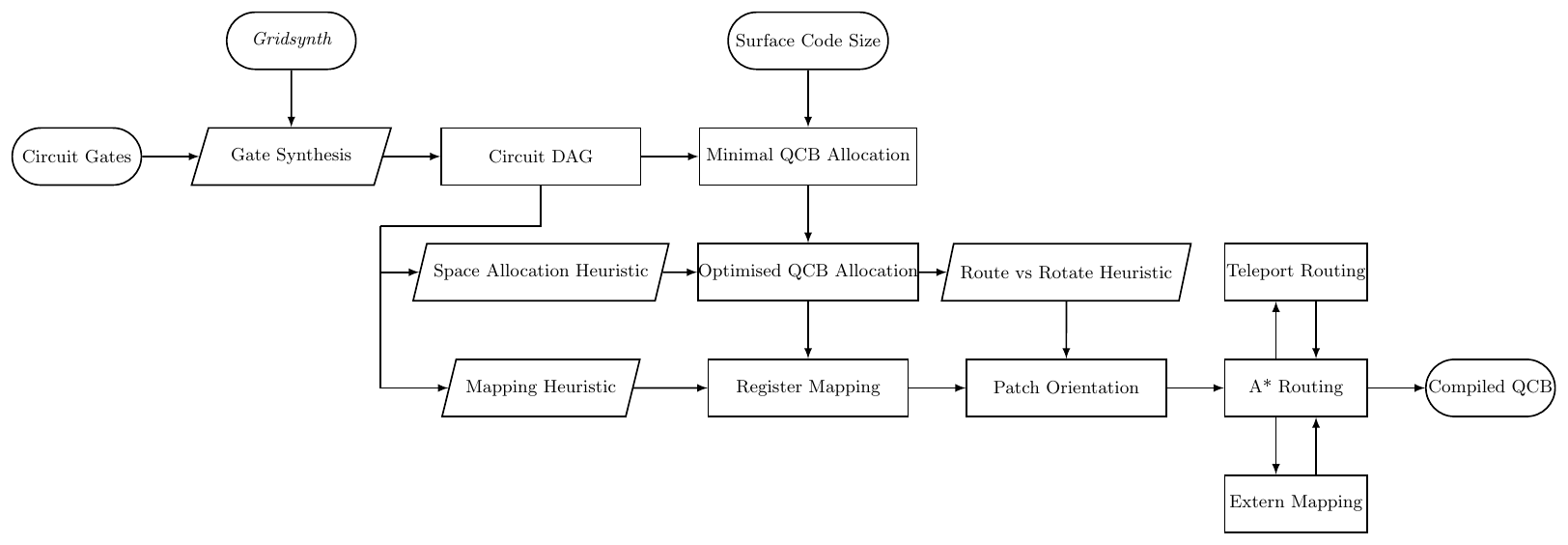}
\caption{Compiler process including parsing, allocation, mapping and routing components.}
\end{subfigure}
\begin{subfigure}{0.47\textwidth}
\begin{center}
    \includegraphics[scale=0.62]{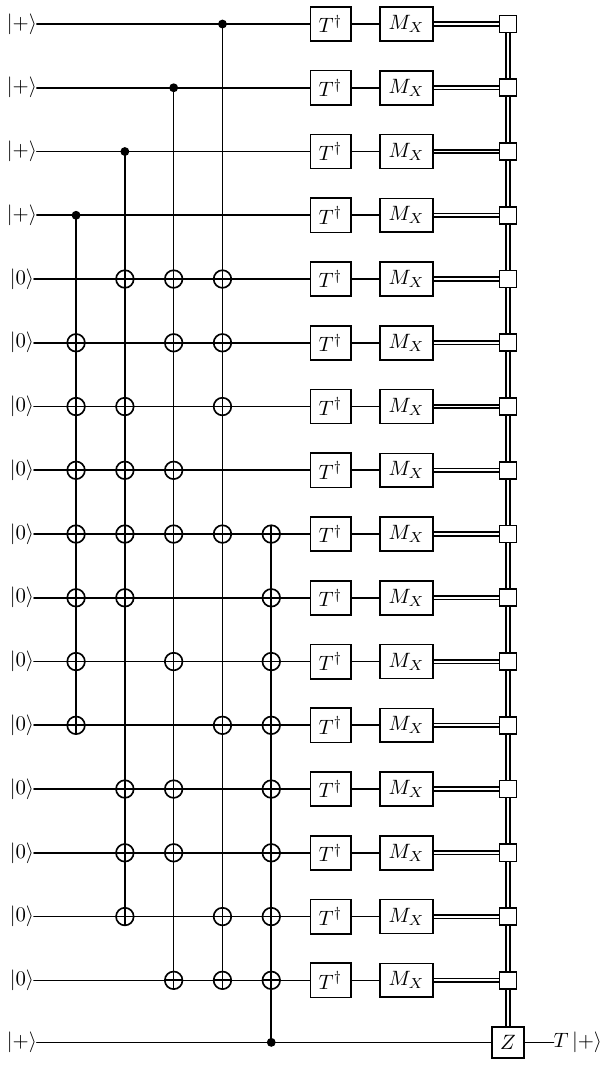}
    \caption{T Gate Factory circuit.}
    \label{fig:t_factory_circuit}
\end{center}
\end{subfigure}
\begin{subfigure}{0.47\textwidth}
\center
   \begin{tikzpicture}
\draw (0, 0) node[inner sep=0] {\includegraphics[]{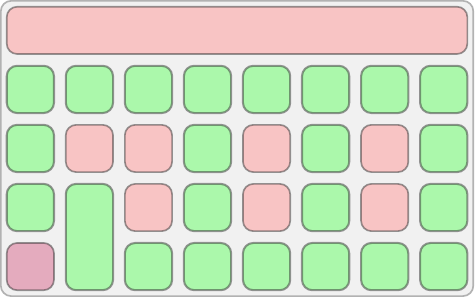}};
\draw (0, 0) node[inner sep=0] {\includegraphics[page=6]{img/results/msf/msf_1}};
   \end{tikzpicture}
\vspace{0.6cm}
\includegraphics[scale=1]{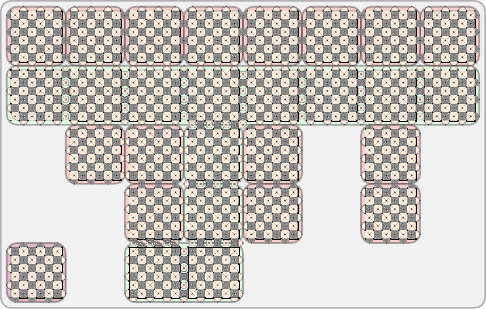}

\caption{Lattice Surgery Target}
\end{subfigure}

\caption{
Overview of the compiler process.
(a) Compiler Pipeline
(b) 15-1 T factory circuit 
(c) QCB allocation for a distillery implementing the circuit in (b), this shows the dependency structure while executing the third CXX gate from that figure. (c) lower Sample lattice surgery operations for one cycle during the execution of (b), this step is preparing the ancillae state required to implement the third CXX gate.   
}
\label{fig:flowchart}
\end{figure*}

\subsection{Outline of the Paper}
\label{sec:outline}

In \Cref{sec:parsing_and_allocation} we describe our parsing and allocation methodologies for the input programme and the QCB.
In particular we discuss the transformation of the programme to a DAG, and the heuristics used for automatic resource allocation.

In \Cref{sec:mapping} we continue describing out compiler, this time looking at the mapping algorithms used for qubit and resource placement.   
In \Cref{sec:routing} this description continues into the implementation of our routing heuristics.

In \Cref{sec:evaluation} we demonstrate the performance of this compiler against a number of different input circuits.
These results are discussed further in \Cref{sec:discussion}.

\section{Parsing and Allocation}
\label{sec:parsing_and_allocation}

Our compiler accepts an input circuit as a time ordered set of operations.   
The structure of the time ordered dependencies over the set of named registers creates a directed acyclic graph (DAG). 
We wish to extend the set of constraints on the structure of the DAG to incorporate the constraints implicit in a lattice surgery based architecture.
These constraints take the form of a type system, which we explain in \cref{sec:parsing_and_allocation/DAG}.

Alongside this we will additionally attempt to design an allocation on a fixed size collection of surface code patches that provides sufficient support to implement the sequence of operations in the input circuit as a series of lattice surgery operations. 
We describe the implementation of this allocation in \cref{sec:parsing_and_allocation/QCB}.

\subsection{DAG Elements}
\label{sec:parsing_and_allocation/DAG}
Each operation in the DAG may either be a native gate, a macro expansion, or an externally defined closure.   
Operations act over symbolically named registers.
These symbolic registers will later be mapped to physical regions of the surface code. 
At this stage of compilation the yet-to-be-mapped registers preserve the association of operations and operands.

\subsubsection{Native Gates}
\label{sec:parsing_and_allocation/DAG/LSNI}

Native gates have direct implementations as pre-defined sequences of lattice surgery operations. 
Example implementations of these lattice surgery operations can be seen in \Cref{sec:ls_ops}.

Linguistically we divide operations into whether they are unary operations or binary operations. 
If they are unary operations we further consider whether
or not they require dedicated computational ancillae.
Patches of the surface code that are required to perform boundary deformations are considered to be a computational ancillae and share the same dependency structure.
Binary operations are non-local, and will require potentially both routing and computational ancillae.

For our purposes we consider native gates to be the set of local unary operations \{Prep $\ket{+}$, Prep $\ket{0}$, Measure $X$, Measure $Z$, $X$, $Z$, $H$\} along with the set of non-local ancillae supported unary operations \{$P$, $P^\dag$\}, and non-local ancillae supported binary operations \{CNOT, CZ\}. 
The requirement for ancillary support qubits will be relevant when we are attempting to route between surface code qubits using a bus. 
Each native gate is further described by an integer number of surface code cycles, and whether it acts along X or Z type boundaries of its surface code patch operands.

For the purposes of magic state support we additionally provide the `unsafe' unary local gates \{$T$, $T^\dag$\}.
Similar gates may be defined to act as proxies for any non-fault tolerant single qubit gates in the gateset.

\subsubsection{Macros}
\label{sec:parsing_and_allocation/DAG/macro}

Bare sequences of native gates do not provide a particularly rich description of a circuit. 
We extend our input circuit with `macro' expressions. 

Macro objects perform symbolic substitutions over the set of registers.
A macro takes an ordered set of registers as arguments, and expands to implement a set of operations over those registers. 

\begin{figure}
        \begin{center}
            \includegraphics[scale=1]{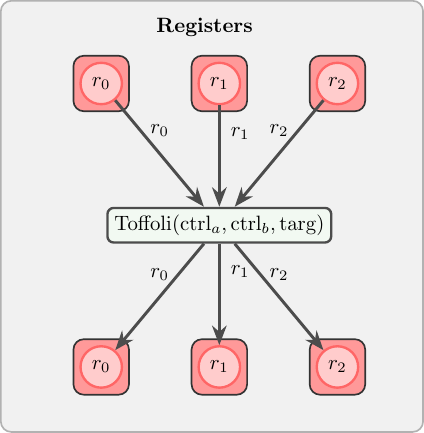}
        \end{center}
        \caption{Toffoli gate expressed as a macro operation.\label{fig:toffoli_macro}} 
\end{figure}

This provides a concise representation of common sequences of operations over varying arguments.  

As an example, a \verb|SH(x)| operation may be defined as a macro. \verb|SH(x) := {H(x); S(x);}|. 
This macro may then be applied to an arbitrary register, which is substituted for \verb|x| when the macro is expanded.   

\subsubsection{Externs}
\label{sec:parsing_and_allocation/DAG/extern}

Universal quantum computation requires non-Clifford gates, which are absent from the surface code native gate set specified above. 
The construction of non-Clifford gates typically requires a magic state factory.

Unlike native gates, which act in place over the space of registers, $T$ gates depending on magic state factories introduce an implicit, temporary dependency on a magic state factory.   
There exist families of magic state factories, and the selection of a particular factory should be abstracted from the construction of the circuit.    
This object is an externally specified dependency and is termed an `extern'.
We distinguish between extern operations and extern objects.
Extern objects are externally compiled QCBs that are described by a height and a width (specifying their memory footprint) and a set of extern operations. 
Extern operations are operations that depend on an extern object, they specify a set of symbolic input and output arguments, along with a number of cycles required to execute that extern operation on the QCB.

In the case of the $T$ gate the gate itself is treated as a local operation that depends on the output of a $\ket{T}$ factory.   
The factory extern object has an explicit lifetime, and once it has been released its memory may be reused.  The $\ket{T}$ factory is allocated and run prior to the $T$ gate.
The gate consumes the output of the factory's $\ket{T}$ operation, and then the factory object is freed, potentially for re-use as another factory extern on which another $T$ gate depends.
Extending from this single factory example we may have simultaneous $T$ gates, requiring distinct and identical copies of factories operating in parallel.

The end goal of our compilation process is to transform an input circuit into an extern operation. 
This compilation unit may then be incorporated into larger programs without repeating the allocation and routing calculations, albeit with an increased runtime cost incurred from passing and returning data to the extern. 
For example an-$n$ qubit adder circuit might be compiled as an extern with a defined height and width, then called repeatedly within a larger circuit.

Lastly, we could consider an accumulator extern. This extern contains register states that persist between calls to the extern. As an example, a set of registers within the extern might be initialised to the zero state, and on each call to the extern these registers are incremented. The extern may then have one function that operates normally over the memory, and another that returns the current state of the accumulator registers.   

By analogy to modern programming languages, these extern objects may act as functions, closures, or classes depending on the context of their construction and invocation.

To distinguish between instances of externs of the same type, symbolic matching is performed on the `name' of the extern to admit the instantiation of an extern.
Reference matching is performed to ensure that operations targeting the same instance of an extern are resolved correctly.

Once a factory is complete a \verb|RESET| operation indicates to the compiler that the unit of memory required by the factory may be deallocated for re-use.
If due to entanglement concerns a factory cannot be deallocated then no \verb|RESET| operation is included and the factory will persist in memory until the termination of the program.

One problem that arises with externs is if a DAG implies the existence of more extern elements than have been allocated, for example, if an operation depends on two $\ket{T}$ resources.
In this case, if only one factory was allocated, this would create a deadlock.
The operation blocks the associated \verb|RESET| operations of the associated factories, while the factory cannot be released to produce another $\ket{T}$ state until the \verb|RESET| operation is called.
To avoid this situation we require that each operation depends on at most one extern.     
To implement a gate that depends on multiple extern resources we instead require that they are buffered by registers in the current scope.

\subsubsection*{Gate Synthesis}

While quantum operations may consist of an arbitrary unitary operation,
only a subset of operations are supported natively by the surface code. 
Other gates must be constructed or approximated by a sequence of gates in some basis set.
We utilise {\it GridSynth} as an efficient gate synthesis stage\cite{gridsynth}
to reduce the input unitaries into a set of legal gates for the surface code architecture. 

We decompose larger programs for the surface code into the operations
defined in \cref{sec:parsing_and_allocation/DAG/LSNI} along with a sequence of $R_Z(\theta)$ gates.
These Z rotation gates are passed to GridSynth with an appropriate approximation precision.

The output of GridSynth is expressed as a sequence of Pauli operations,
Clifford operations and $T$ gates.
These $T$ gates are in turn expressed as dependencies of extern operations. 

Angular rotations that are smaller than the precision provided are a programmer concern.
If the identity is within $\epsilon$ of the requested gate Gridsynth will report an empty gate sequence.

\subsection{Patch Allocation}
\label{sec:parsing_and_allocation/QCB}

The goal of this section of compilation is to decide upon a placement of registers, route, extern and IO elements for the given QCB.  
In this compiler we will assume the fixed allocation of memory elements to regions on the surface code.

To this end we construct and satisfy an initial definition of a minimal QCB that implements a target circuit.
We describe a greedy incremental placement heuristic, through which we attempt to iteratively allocate additional resources to the QCB to speculatively reduce the cycle count of the lattice surgery implementation of the circuit.
To ensure that a given QCB layout correctly implements a given circuit we establish a set of constraints:
\begin{itemize}
\item All circuit level qubits must be uniquely associated with a surface code patch on the QCB.    
\item It must be possible to implement all two qubit operations using routing ancillae between the operand qubits.    
\item Any circuit qubits that perform operations that would require surface code ancillae to implement must have access to ancillae qubits to complete those operations.  
\item Any operations that depend on a resource state must be reachable from a factory that generates that resource.    
\end{itemize}

These functional constrains form the basis of a set of placement constraints, which informs the overall implementation of the placement algorithm.   

\subsubsection{QCB Patch Placement Constraints}
   
For a successful allocation, we must ensure that the final allocation of patches of the surface code admits all operations specified in the circuit.
This can be fulfilled by the following conditions: (1) all multi-qubit operations are supported either locally or non-locally using a routing bus, (2) all operations that require ancillae have access to those ancillae, and (3) all operations that require magic states have access to a magic state factory.

To satisfy these constraints we introduce a type system over surface code patches:

\textbf{Register:} Each circuit-level qubit corresponds to a single `register' type patch.   
For a legal mapping of circuit qubits to the current scope to exist, we now need only ensure that the number of register patches is equal to or greater than the number of qubits in the circuit.

\textbf{Route:} patches provide ancillae support for non-local operations.
This involves the construction of appropriate ancillae states between the target registers, a process termed `routing'. 
Rather than only supporting two qubit operations within the circuit, we instead enforce that the adjacencies of all routing patches in the current scope must form a connected graph.
This routing graph is termed the `bus'.   
Additionally we require that all registers are placed adjacent to a routing patch, such that all registers are connected to the bus.  
From these constraints we ensure that all possible multi-qubit operations are supported, and hence all multi-qubit operations in the circuit are supported.   

As some unary operations also require the use of ancillary qubits, we similarly allow the routing patches to be used as these ancillae.       
We will typically allocate registers in bulk regions. As a result we enforce the stricter criteria that registers must be connected to the bus from above or below.

As an addendum, at some stages during compilation patches may be marked as local routing patches.  
These segments are unallocated as either externs or routes and are not connected to the routing bus. 
They are effectively reserved 

\textbf{Extern:} To support arbitrary resource states we define an `extern' type patch.
Extern type patches must be members of contiguous extern regions on the QCB with defined heights and widths.
Extern operations may be associated with a patch if the height and width required by the operation is equal to or less than that of the patch.  
To ensure that these extern resources may be routed to registers in the current scope we require that the bottom edge of the extern region is completely connected to the bus.
When an extern is allocated, it's bottom edge supports ordered inputs and outputs from left to right.

\textbf{IO:} Finally, we define an IO type, this is placed on the bottom edge of the QCB and otherwise acts as a register.  
Circuit qubits that are also inputs or outputs to the current QCB are mapped to IO patches rather than register patches. 
As with registers IO elements must be connected to the bus, but as their bottom edge is at the edge of the QCB they must be connected from above.   
As externs are connected to the bus along their bottom edge, this implies that IO patches are connected to their internal bus from above, and the bus of the calling scope from below.

\begin{figure}
    \begin{centering}
    \begin{subfigure}[b]{0.5\textwidth}
        \center
            \includegraphics[scale=0.4]{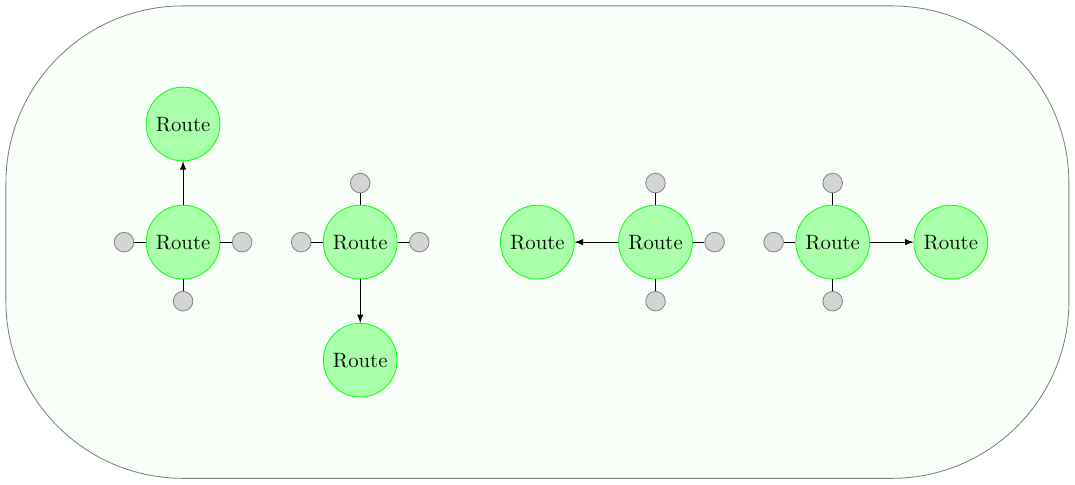}
    \end{subfigure}
    \end{centering}
\vspace{-0.4cm}
\begin{center}
    \begin{subfigure}[b]{0.2\textwidth}
    \includegraphics[scale=0.4]{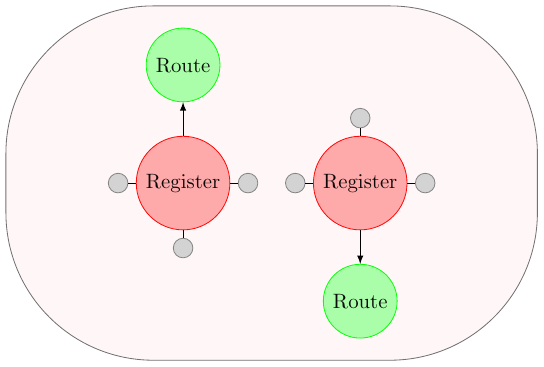}
    \end{subfigure}
    \begin{subfigure}[b]{0.135\textwidth}
    \includegraphics[scale=0.4]{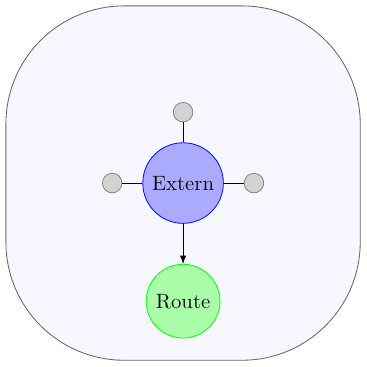}
    \end{subfigure}
    \begin{subfigure}[b]{0.12\textwidth}
    \includegraphics[scale=0.4]{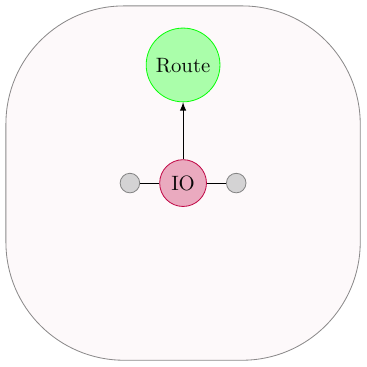}
    \end{subfigure}
\end{center}

    \caption{
Allocation constraints for Routes, Registers, IO and Extern patches.
 The placement of each surface code patch on the QCB must satisfy at least one of these constraints.}
    \label{fig:allocation_constraints}
\end{figure}

These placement constraints are summarised in \cref{fig:allocation_constraints}.

With the exception of the initial placement of an element on the QCB, these constraints may be treated as a set of placement invariants.   
As a base case, we make an initial placement of a register or extern element with an associated bus consisting of routing nodes.
All subsequent placements of externs or registers must be adjacent to a route node, and that route node must be adjacent to an already placed route node.  
By induction, each subsequent placement is connected to the existing bus.

\subsubsection{Placement}

The full set of placement rules are detailed in \Cref{sec:register_placement}. 

In general, we first place an extern or register element with associated routing nodes in the top left corner of the QCB.
This initial route forms the first element of our bus.   
IO elements are placed along the bottom left of the QCB, with an associated set of routing nodes that are not flagged as being connected to the bus. 
Subsequent placements of extern or register nodes are associated with further routing nodes that must be connected to the existing bus. 
Connectivity to the IO bus is deferred from consideration until it is connected to the existing bus.

At each stage of the placement process there is a choice as to the placement of either an extern, or a register element.   
Register placements ensure that the leftmost element of the register is a route that is connected to the bus. 
Similarly extern placements ensure that the row below an extern placement is a route connected to the bus.

If a placement would connect the IO routes with the bus we may cease fretting about this unconnected route.
Alternatively, as the only placements are register or IO placements, the bottom edge of an extern placement is a route element, while the left element of a register placement is a route element.   
As our first placement was an extern or register element in the top left corner of the QCB with the initial bus directly below it we now have established an invariant for connecting the IO to the bus.     

By placing a route directly along the left edge of the QCB from the IO we will either encounter the left edge of a register, the bottom edge of an extern, or our initial placement. 
In all three cases this represents a routing element that is already connected to the bulk of the bus.

\subsection{ DAG Evaluation Heuristic}

To support decision making by the allocator, we require a heuristic that takes a QCB and a DAG and evaluates a figure of merit.   

Given the resource constraints imposed by the finite memory of the QCB, we expect one of the dominant sources of space-time volume costs to be idling costs.   
These idling costs are borne out in the runtime of the DAG on a given QCB, which we have selected as our heuristic figure of merit. 
The goal of this evaluation heurstic is then to determine an approximate cycle count for a given DAG and QCB pair.

Each lattice surgery operation takes a fixed integer number of cycles to implement.
The figure of merit for the heuristic is the total number of cycles required to implement all gates in the DAG.
Extern dependencies require a free unit of sufficiently sized memory as an additional dependency.
Non-local operations require a routing channel as an additional dependency.

\begin{figure}
\center
            \includegraphics[width=0.48\textwidth]{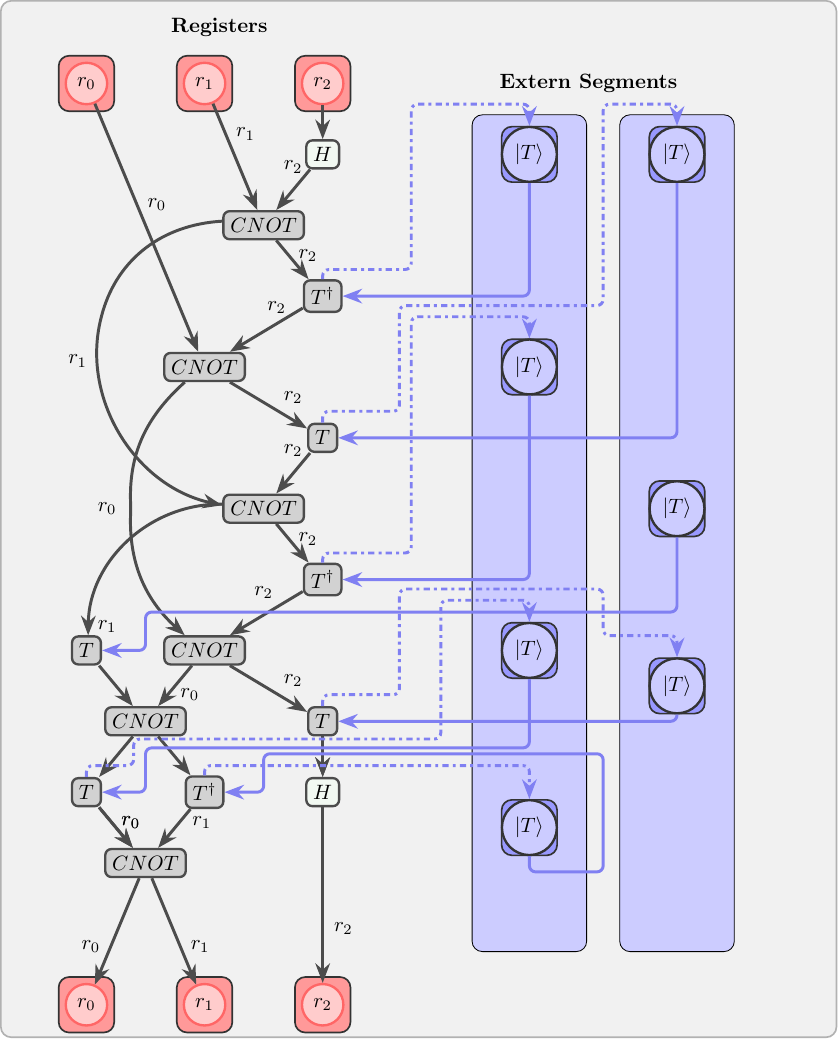}
        \caption{\small
Example DAG implementing a Toffoli gate with implicit dependencies enforcing a constraint on the number of allocated extern segments that are (in this case) producing $\ket{T}$ resource states as dependencies of externally defined $T$ gates.
Dark grey operations indicate non-local gates which compete for an integer number of routing resources.
This integer is an input to the heuristic and is a proxy for the number of routing channels avaliable on the QCB. 
Dashed lines indicate barrier based dependencies on the allocation of each of the $T$ factory objects.
These dependencies prevent extern allocation sequences that would result in deadlocks.   
\label{fig:toffoli_dag_thresh}
}
\end{figure}

Our heuristic takes a set of memory elements explicitly flagged for extern and factory use, along with an integer number of routing channels.
The number of simultaneous non-local operations is bounded by the number of routing channels.
The number of simultaneous factory and extern operations is bounded by the number of memory regions that satisfy those operations, and whether they are already allocated to an operation. An example of this extern ordering can be seen in \ref{fig:toffoli_dag_thresh}.

Our first problem with this model is that factory style resources have no dependencies.
By pre-allocating factory operations in the DAG to individual gates we risk a situation where the physical extern patches are allocated to factories that depend on unresolved and unallocatable factories from earlier in the circuit, forming a deadlock scenario.   
To prevent this situation we construct barriers that prevent the allocation of externs until all dependent gates have been satisfied.
This is equivalent to introducing a dependency to the factory onto the gates in the barrier. 
Similarly the shared contention for the routing resources may be modelled as a semaphore.

As gates in the DAG compete for mutually exclusive resources, we construct a priority scheme for scheduling resources to operations.  
Each operation is assigned a `slack' value given by the difference between the earliest time for the execution of that operation and of any dependent operations.
This represents the number of cycles which a gate may be delayed without increasing the overall circuit depth.  
Operations with multiple dependencies which are only resolved much later in the program have higher slack values, and indicate that the scheduling of that particular gate is not urgent.

Operations are ordered by their slack value. 
The heuristic attempts to schedule operations in this order.  
If two operations share a slack value they are attempted in the order in which they appear in the original DAG sequence.

\begin{figure*}
\center
    \begin{subfigure}[b]{0.16\textwidth}
\includegraphics[page=1, scale=0.14]{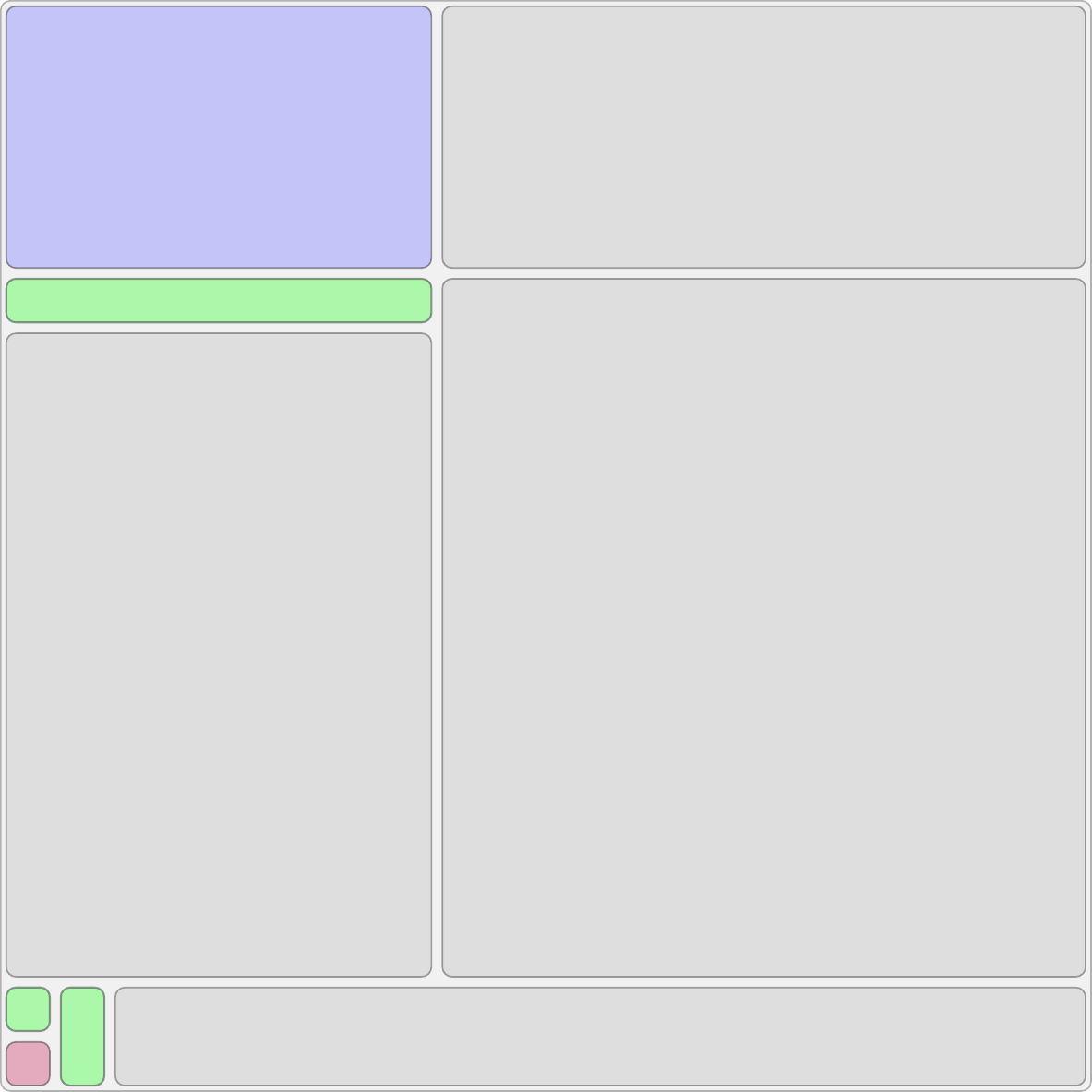}
    \end{subfigure}
    \begin{subfigure}[b]{0.16\textwidth}
\includegraphics[page=2, scale=0.14]{img/allocation_walkthrough/t_nest_alloc}
    \end{subfigure}
    \begin{subfigure}[b]{0.16\textwidth}
\includegraphics[page=3, scale=0.14]{img/allocation_walkthrough/t_nest_alloc}
    \end{subfigure}
    \begin{subfigure}[b]{0.16\textwidth}
\includegraphics[page=4, scale=0.14]{img/allocation_walkthrough/t_nest_alloc}
    \end{subfigure}
    \begin{subfigure}[b]{0.16\textwidth}
\includegraphics[page=6, scale=0.14]{img/allocation_walkthrough/t_nest_alloc}
    \end{subfigure}
    \begin{subfigure}[b]{0.16\textwidth}
\includegraphics[page=29, scale=0.14]{img/allocation_walkthrough/t_nest_alloc}
    \end{subfigure}
\caption{\small
Example QCB construction for a nested/concatenated T distillery.
The initial placement includes an IO element for the output $\ket{T}$ state, along with an extern of sufficient size to accommodate a $T$ factory.
Subsequent placements (the second and third stages) ensure that a sufficient number of registers have been allocated.  
At this point the initial correctness guarantees have been met, and the optimisation phase begins.
The heuristic finds that the main runtime constraint is the lack of additional $T$ factories, and begins allocating more $T$ factories to the QCB.    
Finally, there are no more legal placements that satisfy the routing constraints, and the remainder of the QCB is divvied up with spaced register elements.    
\label{fig:allocation_walkthrough}
}
\end{figure*}

\section{Mapping}
\label{sec:mapping}

Following successful completion of the tasks in \cref{sec:parsing_and_allocation},
we have allocated surface code patches to be used for exactly one of: 
store logical quantum information,
transmit logical quantum information as part of a bus, or
mark for use as part of a separately defined extern element.
The purpose of this section is to explain how logical qubits
will be mapped to storage patches.

This mapping has two stages:
register mapping (\cref{sec:mapping/register}) and
orientation tracking (\cref{sec:mapping/orientation}).
The overall idea of our mapping heuristic is to construct a 
Steiner tree that connects distinct connected collections of storage patches,
perform round-robin allocation of logical qubits to storage patches,
and finally to identify the preferred strategy
for each logical qubit to be connected to other logical qubits
during the execution of the logical quantum circuit.

\subsection{Register Mapping}
\label{sec:mapping/register}

Having assigned registers we now consider mapping logical qubit labels to physical patches on the device. 

Our mapper will attempt to maximise the bus distance between the allocations of qubits that contend for routing resources in the DAG.  
This allocation seeks to reduce the number of collisions during routing. 

Our goal in this section is to reduce the bus to an approximate Steiner tree, where leaves of the tree represent register, extern and IO elements.
Our mapping algorithm then performs a weighted round robin allocation over the tree.

\begin{figure}
        \begin{center}
   \includegraphics[width=0.8\linewidth]{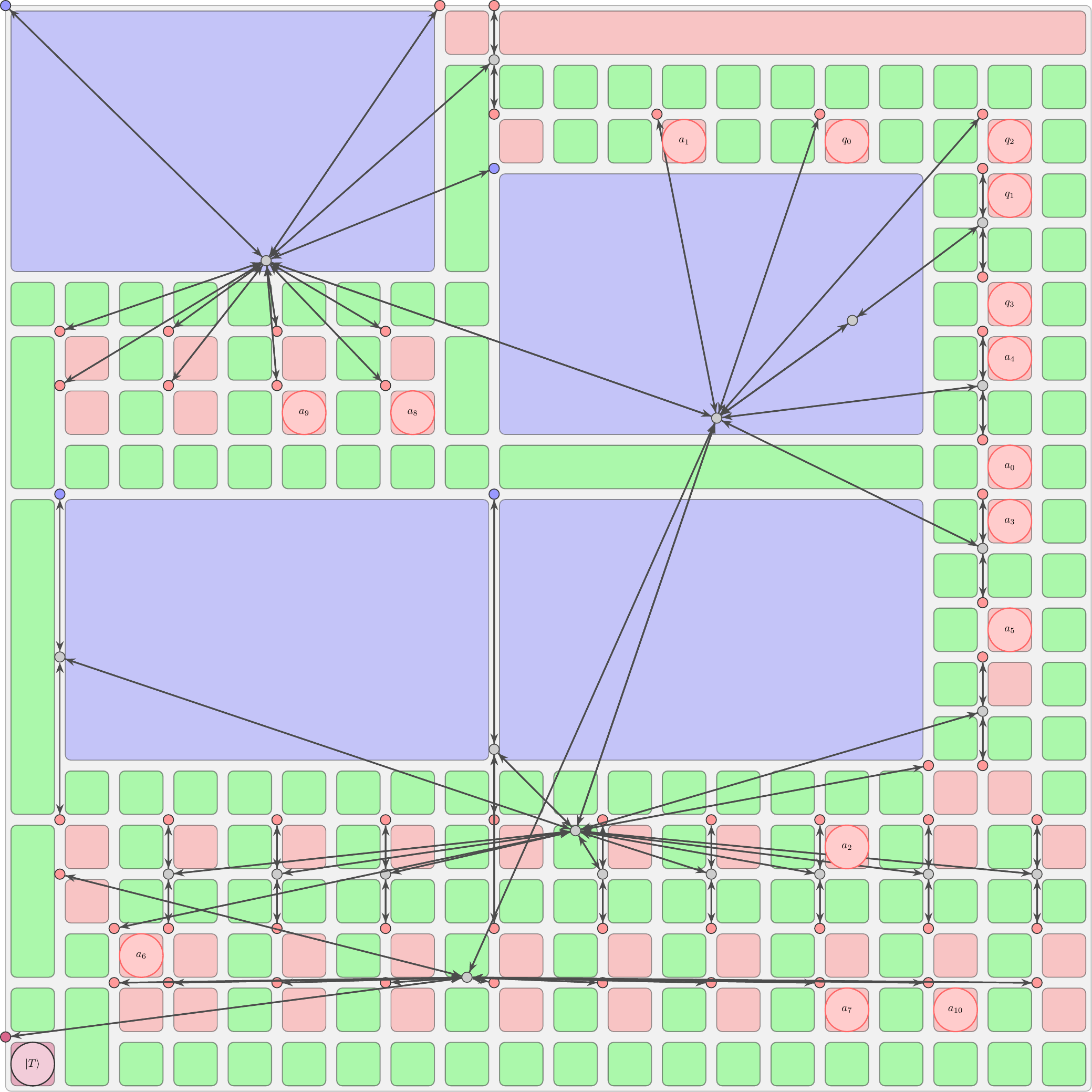}
        \end{center}
        \caption{ \label{fig:intra_register_mapping} Approximate Steiner tree construction for the QCB layout from \Cref{fig:allocation_walkthrough}. Mapped registers are labelled, while unmapped registers will be reallocated as routing ancillae.}
\end{figure}

\subsubsection{QCB to Tree}

In this section we will describe the process of mapping the QCB to an approximate Steiner tree.

From the construction of our routing segments, the bus on the QCB forms a complete graph that connects all registers, externs and IO elements.
We begin by truncating non-branching elements of the routing graph. 
We will use the number of routing nodes as a heuristic for the accessibility of a register, and so are only interested in counting elements that are topologically distinct on the bus graph.           

From this truncated graph representation, we treat all non-routing elements on the graph as root nodes of sub-trees.
As described in \cref{alg:graph_to_tree} each of these subtrees is then grown and merged into a single spanning tree.
Vertices in the tree are weighted by the number of routing edges consumed by the growth of that particular subtree.

Adding a routing edge to a subtree propagates a value $k$ down the sub-tree such that at each vertex with $n$ children the weight of each child is increased by $\frac{k}{n}$. 
At the top level each edge propagates a value of $1$.
The value of all non-leaf nodes is then the maxmimum value of their children.
These values establish an ordering over the children of each node, where the value corresponds to the amount of free routing space around the child.

\begin{algorithm}
\caption{Routing Graph to Tree}
\label{alg:graph_to_tree}
\begin{algorithmic}
    \STATE Let each register $i$ be the root of a subtree $s_i$
    \STATE Let each subtree be associated with a set of nodes $e_i$ initially containing $s_i$ 
    \WHILE{$|S| > 1$}
        \FOR{$s_i \in S$} 
            \STATE Expand $e_i$ to include all routing node neighbours of elements of $e_i$ 
            \STATE If a node is already associated with another set $e_j$, flag it. 
        \ENDFOR
        \FOR{Each flagged node}
            \STATE Create a new root node and associate all subtrees that share a node as children 
            \STATE Remove the children from $S$ and add the new subtree to $S$
            \STATE Unflag the edge 
            \STATE Propagate a value of 1 from the new root node.
            \STATE The value of the new root node is set as the maximum of its children
        \ENDFOR
        \FOR {Each unflagged node that was added to a single subtree this round}
            \STATE Propagate a value of 1 from the root node of the grown subtree
        \ENDFOR 
    \ENDWHILE
\end{algorithmic}
\end{algorithm}

Once the tree is built we use it to recursively allocate register symbols to physical registers on the graph.
Each vertex maintains track of both its score as a function of the maximum of its children's scores, and the total number of allocatable registers for that sub-tree.
Lastly it tracks how many registers have been allocated to each child.

Allocation at all levels of the tree is performed via round robin, with the following initial order:
Each branch will allocate a register symbol to the branch with fewest allocated registers, in order of the highest free routing score.  
This ensures that subsequent allocations are placed on registers that are distant on the bus graph.

We order the register symbols in the DAG by the number of routing contentions.
Taking the most-contended register, we then sort the other registers it contends with by the number of contentions.
These registers are then allocated in that order, with the goal of minimising the maximum number of routing contentions encountered in the DAG.  
This process is then repeated until all registers have been allocated.  
If at any point a vertex no longer contains any allocatable registers then it is pruned from the tree.
As the generated QCB guarantees that number of physical registers equals or exceed the number of DAG registers there will always be a sufficient number of registers to complete this allocation.

Our earlier register placement deposited horizontal lines of registers across the QCB.  
While the current mapping approach allocates circuit qubits to register objects, it does not delineate which surface code patch within a register the qubit is associated with.   
To perform this secondary mapping we attempt to maximise the number of free edges within the register block.  
Without reference to the current state of allocated ancillae, this minimises the number of lattice surgery patch rotations, while also providing ancillae patches that do not contend with existing route ancillae.    
To decide on the next placement we take the largest unallocated region of the register patch and perform a mapping such that this region is split.         
An example of this placement strategy can be seen in \Cref{fig:intra_register_mapping}.

\begin{figure}
        \begin{center}
            \includegraphics[scale=0.9]{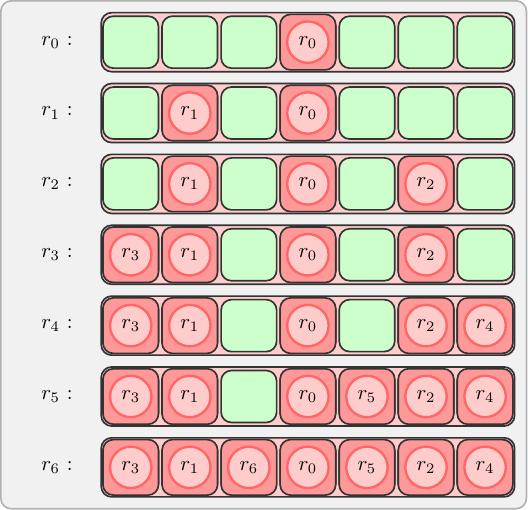}
        \end{center}
        \caption{ \label{fig:intra_register_mapping} Greedy mapping within a register block to preserve exposed edges.} 
\end{figure}

\subsubsection{QCB Cleanup}

Having allocated the logical qubits to physical patches we are left with cleaning up the remainder of the QCB.
For this we have two considerations, the first being unallocated register segments, and the second are the local route segments.

Unallocated register segments are re-labelled as routing segments.
As register segments must already be connected to routing elements any routing segments created in this manner are already connected to the routing network.
If any local routing patches are adjacent to a routing patch then that local routing patch and all connected local routing patches are replaced with routes. 
This finalises the placement of the elements on the QCB, but does not yet associate extern regions with DAG elements.

\subsection{Orientation Tracking}
\label{sec:mapping/orientation}

The earlier register placement requirement does not necessarily expose both an X and a Z boundary of the surface code patch.
The next pass is to determine whether routing to or rotating is a preferred strategy for each patch.

For each allocated register we check if both an X and a Z type boundary are connected to the routing bus. If this is the case then routing is the preferred strategy, otherwise patch rotation is preferred.
When operations are performed on the QCB rotation instructions are injected into the DAG for the appropriate qubits.   

\section{Routing}
\label{sec:routing}

Having constructed the QCB and assigned the register elements routing between patches on the QCB is performed using $A^*$.
As with the heuristic evaluation operations are ordered by their slack in the DAG and operations with lower slack are routed first.
Once an operation is routed the routing ancillae patches required to implement that operation are locked for the duration of the operation.

\subsection{Edge Tracking}
The instructions for each operation describe a set of dependencies over the edges of each operand patch. 
As a first pass before routing we assess each allocated qubit and tag whether it is possible to route independently to each edge, or whether a rotation operation is required.   

When the register patch participates in an operation, if the required edge is already exposed to an unused routing ancillae then no changes are required.
If this is not the case then if the patch is tagged as requiring a rotation, a rotation operation is injected into the DAG, otherwise the gate is delayed until the appropriate edge is accessible.

\subsection{Extern Allocation}
While the register and IO patches have fixed locations, this is not true of the extern allocations. We present three different extern allocation methods. 

The first uses the allocation of the externs from the heuristic evaluation of the DAG. When an extern is allocated to a sequence of gates it is locked until a \verb|RESET| instruction is encountered. If another gate would require the same physical extern for a distinct allocation then it blocks until the extern is released.       

The second allocation is a FIFO queue over types of externs. Each extern segment is bound to a particular type of extern gate. When that gate is encountered it is passed to the FIFO queue. Once a physical extern segment matching that is bound to that type of gate is free the first element of the queue is bound to the extern.    

The final allocation method is to share all extern segments between all queues. Extern allocation is validated against the height and width of each physical segment. 
Once a gate has been associated with a physical extern the IO channel of the extern describes which patch along the bottom edge of the extern segment is associated with each IO element. 
If all externs share a type then this method is equivalent to the FIFO queue.

\subsection{Disjoint Path Routing}

A recent technique in alleviating route contention for lattice surgery is to measure out intermediary ancillae\cite{Beverland_2022, padda2023improving, devulapalli2022quantum}. 
We implement a greedy approximate disjoint path routing as a compilation option, using a construction similar to the one shown in \ref{fig:sc_cnot_circ} .

Once a route has been established  a check is performed on all non-terminating ancillae (those that are not adjacent to one of the operands).
For each of these ancillae if their last used cycle was more than two cycles prior they are listed as a candidate disjoint path.
If at least three adjacent ancillae are candidates then they are back-propagated.

These disjoint path routes exhibit numerous trade-offs when compared to direct routing.    
Non-local multi-target operations scale with a time over head of $O(\log(s))$ in the number of measurement rounds to decoder rounds where $s$ is the distance of the path, and with a spatial overhead of  $O(s\log(s))$\cite{latticesurgerywithatwist}.
In our construction the additional spatial overhead is absorbed into the patch sizing.

Disjoint path based route construction requires only constant overheads and parallelisable nearest neighbour interactions.
Conversely, the number of cycles required to construct the bell state paths is doubled.
Holding ancillae qubits for multiple cycles additionally increases the space-time volume of the circuit in exchange for potentially freeing up ancillae in the current cycle and reducing blocking on routing ancillae.

The benefits of disjoint path routing then depend on the length of the paths that they replace, and on the potential routing delays that they alleviate. 

\subsection{Compiled QCBs are Externs}

Having provided a sequence of routing operations required to implement the circuit, and determined the total number of cycles required to implement that circuit for a QCB of fixed height and width, with a fixed IO channel, the compiled QCB may now itself be treated as an extern element.    

Each compiled element is now a predicate for an extern in another program.  
This provides our primitive for the recursive composition of compiled lattice surgery blocks.

\section{Evaluation}
\label{sec:evaluation}

In this section we will evaluate the performance of our lattice surgery compiler against a collection of common quantum algorithm primitives.

In all cases, we evaluate our compiler by determining the space-time volume and the number of surface code cycles (`tocs') required to implement an operation with given pre-compiled externs under the spatial constraints set by the QCB size.

\subsection{Basic Logical Operations}
\label{sec:evaluation/logic_gates}

\subsubsection{Random CNOT Network}
\label{sec:evaluation/logic_gates/cnot_volume}

For our first benchmark we will consider input circuits comprised of random CNOT operations.
CNOT operations are mediated using a collection of routing ancillae as described by the circuit in \Cref{fig:sc_cnot_circ}.
The CNOT controls and targets are selected by randomly permuting the qubits such that each round of CNOT operations incorporates one CNOT gate involving each register. 

These circuits require no magic state resources, and all have constant circuit depths.
In the architecture-agnostic case the runtime of these circuits would be constant, and the space time volume would be a constant multiple of the number of CNOT gates performed. 

\begin{figure}
  \begin{subfigure}{0.45\textwidth}
    \begin{center}
      \includegraphics[scale=0.6, trim={0 0 0 0}]{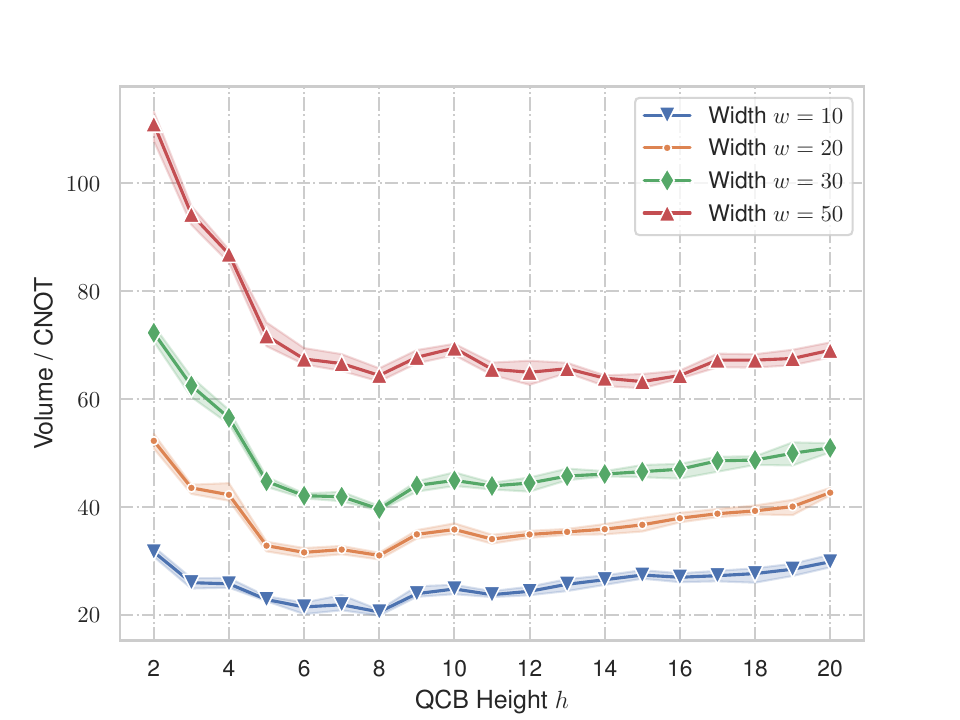}
    \caption{\small Average Space Time Volume for each CNOT\label{fig:cnot_volume}}
    \end{center}
  \end{subfigure}
  \begin{subfigure}{0.45\textwidth}
    \begin{center}
        \includegraphics[scale=0.6, trim={0 0 0 0}]{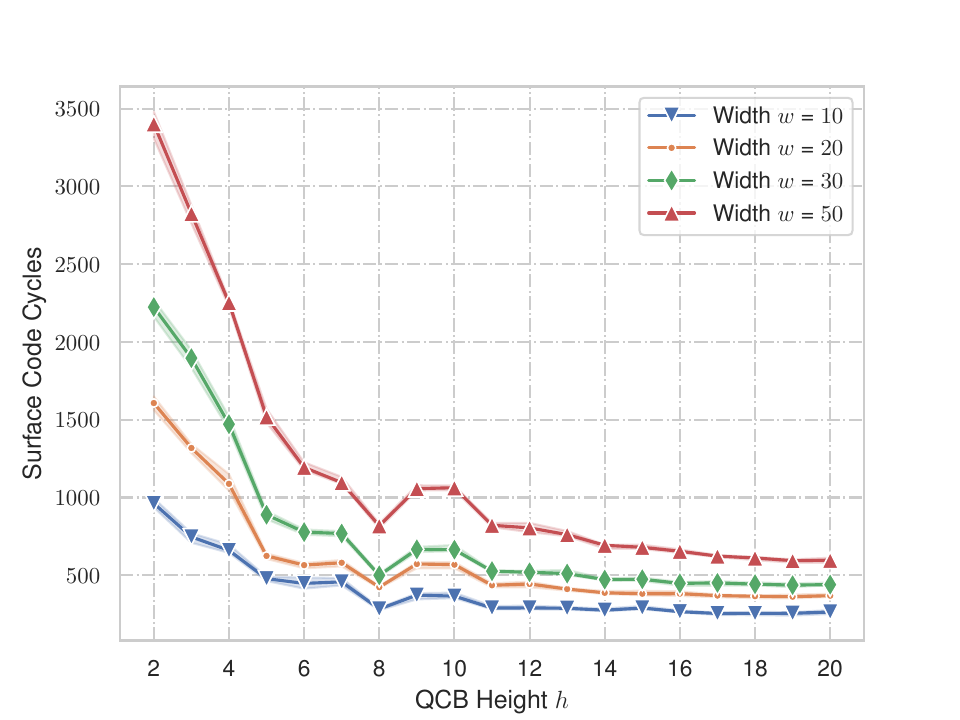}
      \caption{\small Runtime of random CNOT circuits. \label{fig:cnot_runtime}}
    \end{center}
  \end{subfigure}
    \caption{\small  100 round random CNOT benchmark. Each round is comprised of a set of $\frac{w}{2}$ CNOT gates with each qubit participating in one gate in each round for a constant circuit depth of $100$.
The runtimes and space-time volume costs exhibit a tradeoff between idling while contending for limited routing resources, and additional space-time volume costs attributable to larger routes.
    This trade-off is attributable to the sparsity of the register allocation, with idling dominating up to $\frac{1}{8}$ registers to routing ancillae, and routing costs dominating beyond this regime. Conversely the runtime of the circuit continues to improve as register sparsity increases as routing costs are constant in time and so only idling costs impact the runtime of the circuit.  \label{fig:cnot_benchmark}}
\end{figure}

Instead, in \Cref{fig:cnot_volume} and \Cref{fig:cnot_runtime} we see the average space-time volume per CNOT operation, and the number of surface code cycles required to implement this fixed depth circuit as a function of the height and width of the QCB.   

As routing has a constant runtime, the only overhead for the cycle count is contention for routing resources.       
As the register density decreases and the height of the QCB increases, the additional routing ancillae reduces route contention, improving the runtime of the circuit.   

Conversely, space-time volume costs depend both on the idle time of the registers, and the length of the routes. 
Below a register density of $1/8$ idling associated with route contention dominates the space time volume costs.    
Above this fraction we see that space time volume costs increase with the greater register sparsity and the associated increase in route length.

\subsubsection{Multi-Level Factory Benchmarks}
\label{sec:evaluation/logic_gates/MSF}

In this section we compile nested 15-1 $T$ factories with varying footprints. 
Rather than using existing hand-compiled designs, we allow the compiler to construct its own layouts and mappings within the memory region specified.

We may nest this initial magic state factory in a second round of distillation where each of the $T$ gates in the circuit are satisfied by the output of a previous round of distillation\cite{Ding_2018, Krinner_2022}.
Each output then depends on a previously compiled extern.
Allocating larger factory footprints increases the number of externs that may be allocated and permits greater parallelism in $T$ gate operations.
Due to the routing and IO channel limitations the compiler produced factories are larger than hand crafted implementations~\cite{gidney2019flexible, not_as_costly}.
An example of the nested layout can be seen in \Cref{fig:msf_qcb_example}.

\begin{figure}
\begin{center}
\includegraphics[scale=0.55]{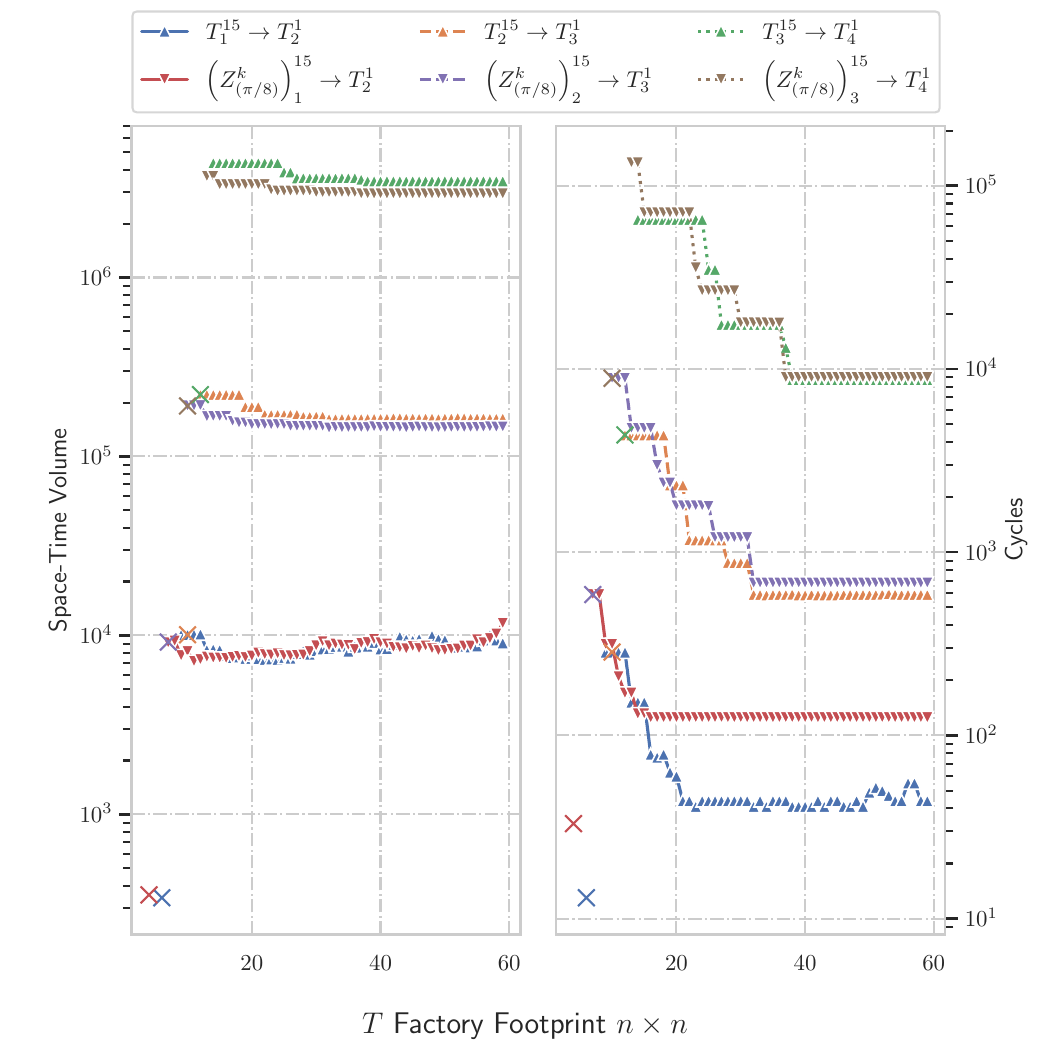};
\end{center}
\caption{Space time volume and cycle times for automatically constructed $T^{15}_{i - 1} \rightarrow T^{1}_{i}$ factories as the number of underlying logical patches varies. For smaller factories the routing costs dominate the reduction in idling costs gained by using a larger QCB, while for larger levels of distillation with longer idling times space-time volume costs may be reduced using larger QCBs with more extern allocations. \label{fig:msf_nest}}
\end{figure}

We consider $T$ distillation using the circuit shown in \Cref{fig:t_factory_circuit}, requiring 15 independent $T$ states, and the slice style circuit from \cite{not_as_costly} which requires 15 $Z^k_(\pi/8)$ gates, each depending on a previously distilled $\ket{T}$ state.       
The main distinction between these implementations is that the demand for $T$ externs in the $Z^k_(\pi/8)$ occurs with each slice, incurring dependencies in the DAG, while the $T^15$ circuit only requires the $\ket{T}$ states in parallel at the end of the circuit, but also require additional ancillae.  
The space-time volume costs and run-times of the first three distilleries of each type can be see in \Cref{fig:msf_nest}.

As with the CNOT benchmark as we increase the size of the QCB we see an initial drop in space-time volume costs attributable to a reduction in idling costs, before the additional routing lengths dominate. 
This exhibits as similar Ahmdahl's law influenced costing, the reduction in space-time volume costs from reduced idling is more pronounced between the allocation of the first and second extern, and is subject to diminishing returns.

The two different $T$ distillation methods exhibit similar space-time volume costs and runtime costs.    
The main advantage of the slice method is that by reducing the number of ancillae the factory fits within a smaller QCB.  
Conversely the greedy allocation heuristic tends to place more externs and achieves better run-times for the non-slice distillation.      

All tiles of the factories operate at a minimum and constant code distance of $d \times d$, as opposed to variable distance tiles.
This produces larger run-times and space-time volume costs than variable tile-size schemes.
In this regime, if a full $15$ factories are placed and the run-time of the factory is upper bounded by the runtime of the nested factories plus a small constant overhead. 

\Cref{fig:msf_qcb_example} shows an example layout of a nested $T$ distillation round as produced as an output by the compiler.

\subsubsection{Toffoli Benchmarks}

Having compiled and benchmarked networks of CNOT gates along with multi-level magic state factories, in this section we combine these elements into a single circuit and compile networks of Toffoli gates. 

We compare three compilation strategies - the first expands Toffoli gates as a DAG containing 7 $T$ gates as shown in \Cref{fig:toffoli_dag_thresh}.
Parallel $T$ gates then contend for factory resources.
The second compiles Toffoli gates as an extern such that the internal operation of each Toffoli occurs in a bounded memory region and that factories are not shared between Toffolis.

In this model, parallel Toffoli gates contend for Toffoli extern resources, and factory operations are delayed until the extern object is allocated.
This modularises and simplifies compilation, while potentially reducing routing costs by ensuring memory locality. 
Conversely, it provides a strict association of $T$ factories to individual Toffoli gates, and prevents the sharing of any additional $T$ states between externs - limiting speedups from factory parallelism when compared to a fully expanded DAG.
The final compiles the Toffoli from a CCZ resource state~\cite{Eastin_2013,Gidney_CCZ}.
This is a midway point between the extern and the factory constructions; the Toffoli gates contend for CCZ resources, while as CCZs are factories they may be executed in advance.

As the fidelity of the Toffoli gates depend on the distillation of non-Clifford states, we recursively compile a range of $T$ and $CCZ$ factories using $\ket{T}^{15}_{i} \rightarrow \ket{T}^{1}_{i+1} $ and $\ket{T}^{8}_{i} -> \ket{CCZ}^{1}_{i + 1}$ distilleries~\cite{not_as_costly,Gidney_CCZ}.  

The $CCZ$ state can be used to perform a Toffoli gate with the addition of local Hadamard gates, while a Toffoli DAG may be constructed from $T$ gates as shown in \Cref{fig:toffoli_dag_thresh}.

\begin{figure}
    \begin{center}
    \includegraphics[scale=0.48]{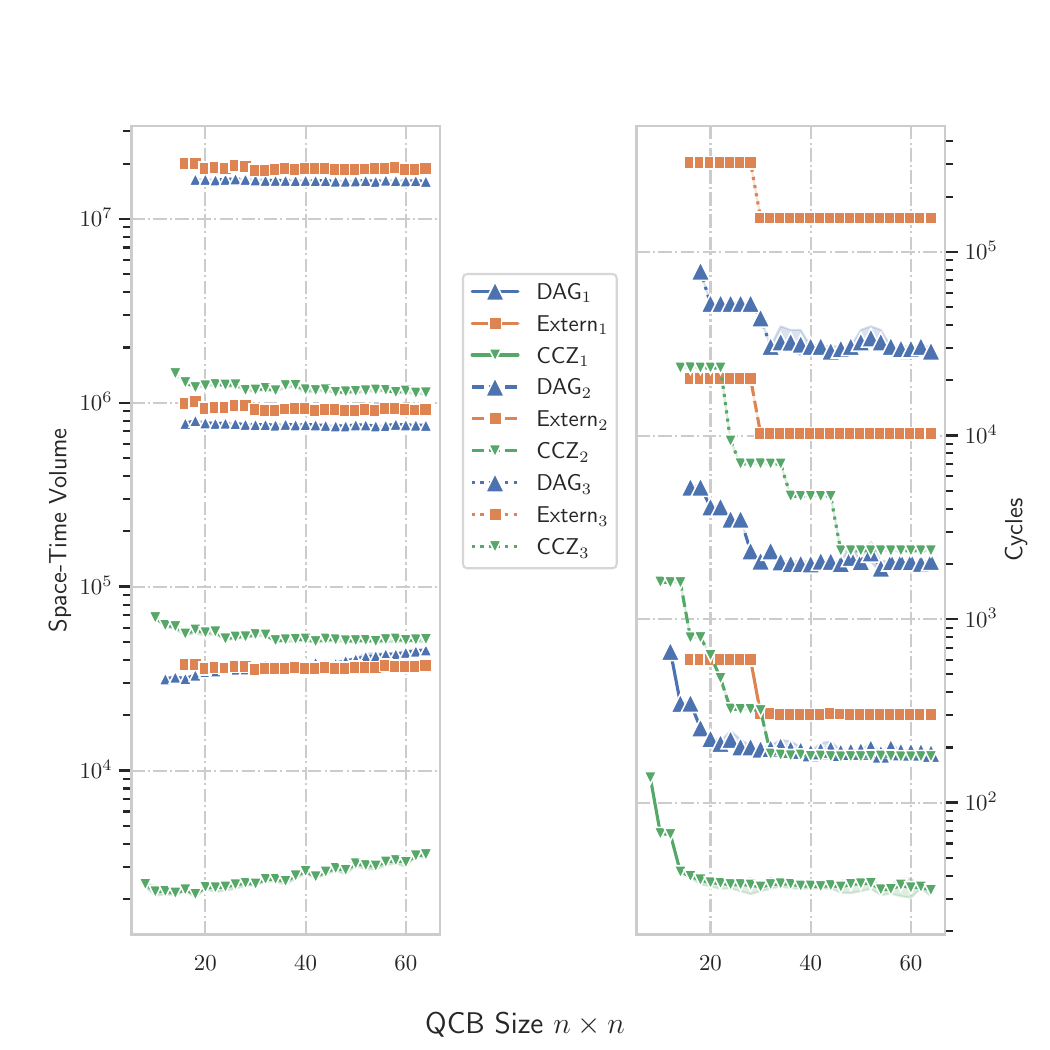}
        \caption{\small Space-time volume costs for circuit implementing 18 Toffoli gates randomly over a set of 6 registers for a range of different Toffoli compilation strategies and distilleries. As the size of the QCB varies the number of $CCZ$ or $T$ factories increases leading to a reduction in the cycle count, while the increased distance between registers leads to a slight increase in routing related space-time volume costs. \label{fig:toffoli_cost}}
    \end{center}
\end{figure}

With the $T$ DAG the network of Toffoli gates is unrolled into a sequence of Clifford operations and factory dependent gates. 
The execution of these factories does not depend on the prior execution of other gates, and may be run in parallel to the remainder of the circuit.
As this dramatically increases the number of gates that are simultaneously scheduled, this increases the complexity of the $A^{*}$ routing and increases compilation times. 
Additionally, this discards the memory locality implicit to the extern models and increases the potential routing space-time volume costs.  
Conversely this allows any Toffoli gate to request any $T$ factory resource, potentially increasing the parallelism of the circuit.

The $CCZ$ factory exhibits a `best of both worlds' approach with parallelism from factory-like behaviour, along with a reduction in the number of attendant gates for compilation.           
Conversely the logical error rate of the $CCZ$ distilleries tends to be higher than the $\ket{T}^{15}_{i} \rightarrow \ket{T}^{1}_{i + 1}$ distilleries\cite{not_as_costly}.

The relative costs of each model can be seen in \Cref{fig:toffoli_cost}.
Minor space-time volume reductions are achieved by the parallelism gained by increasing the size of the QCB, however for smaller factory methods this is dominated by routing overheads.


\subsubsection{Unary Rotations}

A non-Toffoli magic state operation is performed by approximating an arbitrary $Z(\theta)$ gate to a given precision. This is implemented using a sequence of Clifford, Pauli and $T$ gates, with the sequence itself depending upon the choice of $\theta$ and the precision of the approximation. We generate these sequences using `gridsynth'~\cite{gridsynth}.     

For a given approximation error $\varepsilon$  

\begin{figure}
\begin{center}
    \includegraphics[width=0.50\textwidth]{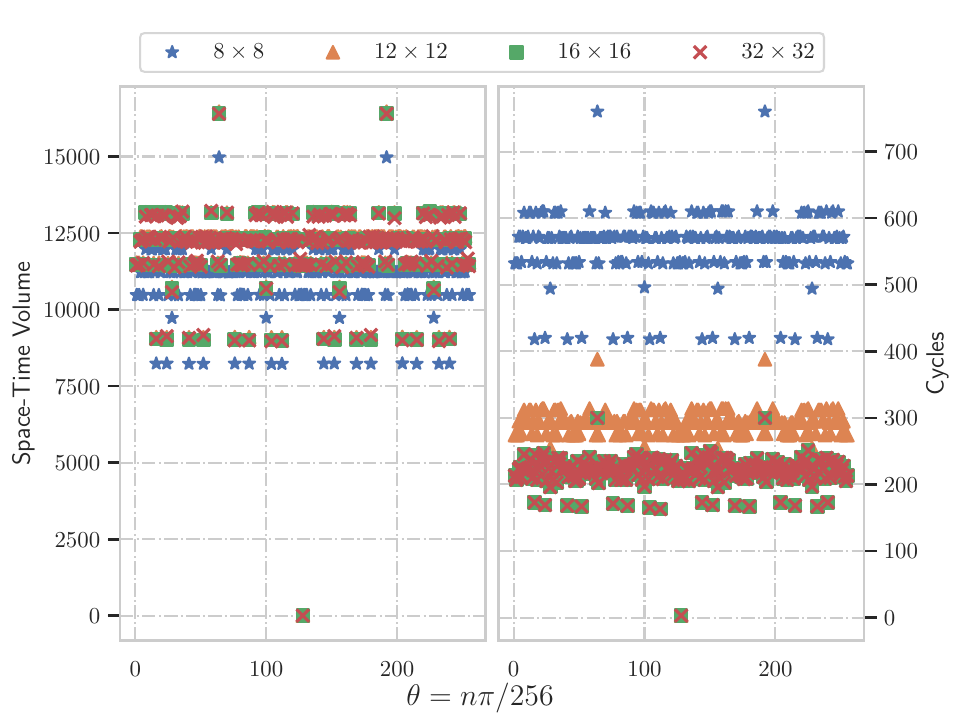}
\end{center}
    \caption{Cost of $Z(\theta)$ rotation where $\theta = \frac{n\pi}{256}$. The drop in the center of the graph is the $S$ gate, which requires no $T$ factories. \label{fig:z_theta_results}}
\end{figure}

In \Cref{fig:z_theta_results} we see the space-time volume and cycles of differing rotation angles for 10 bits of precision. 
The dips at the centre of the plot correspond to $S$ gates that require no $T$ factories.  
As with the
Of note, is that based on the runtime of the factory that there exists a constant number of factories that saturates the capacity for feeding $T$ gates onto a register as part of an $R_z(\theta)$ rotation. 
This behaviour is noted where the $16\times$ and $32\times$ factories exhibit the same runtime.  
As with our other small factory models, the reduction in runtime and associated idling overheads does not supplant the increase in routing costs associated with a larger QCB.

\subsection{Quantum Algorithm Primitives}
\label{sec:evaluation/primitives}

\subsubsection{Multi-Controlled X}

Using the Toffoli gate as a primitive, we may extend our benchmarking to general multi-controlled $X$ gates. 

We opt for an implementation that constructs a multi-controlled Toffoli as a chain Toffoli gates, we do not use the `elbow trick' for these figures.
For larger $C^nX$ gates we can construct $C^aX$ and $C^bX$ such that $a + b = n$, perform a Toffoli gate between the outputs and uncompute the ancillae.  
This is more expensive than a single Toffoli slide, especially when we uncompute the ancillae between each invocation of the extern, but is parallelisable and is faster to compile. 

These QCBs may be compiled either using Toffoli gates as an extern, and hence preventing the sharing of magic states between extern regions, or by using magic state factories as externs, thereby increasing the number of extern allocations, and the number of gates in the circuit. 

\begin{figure}
\begin{center}
\includegraphics[scale=0.45]{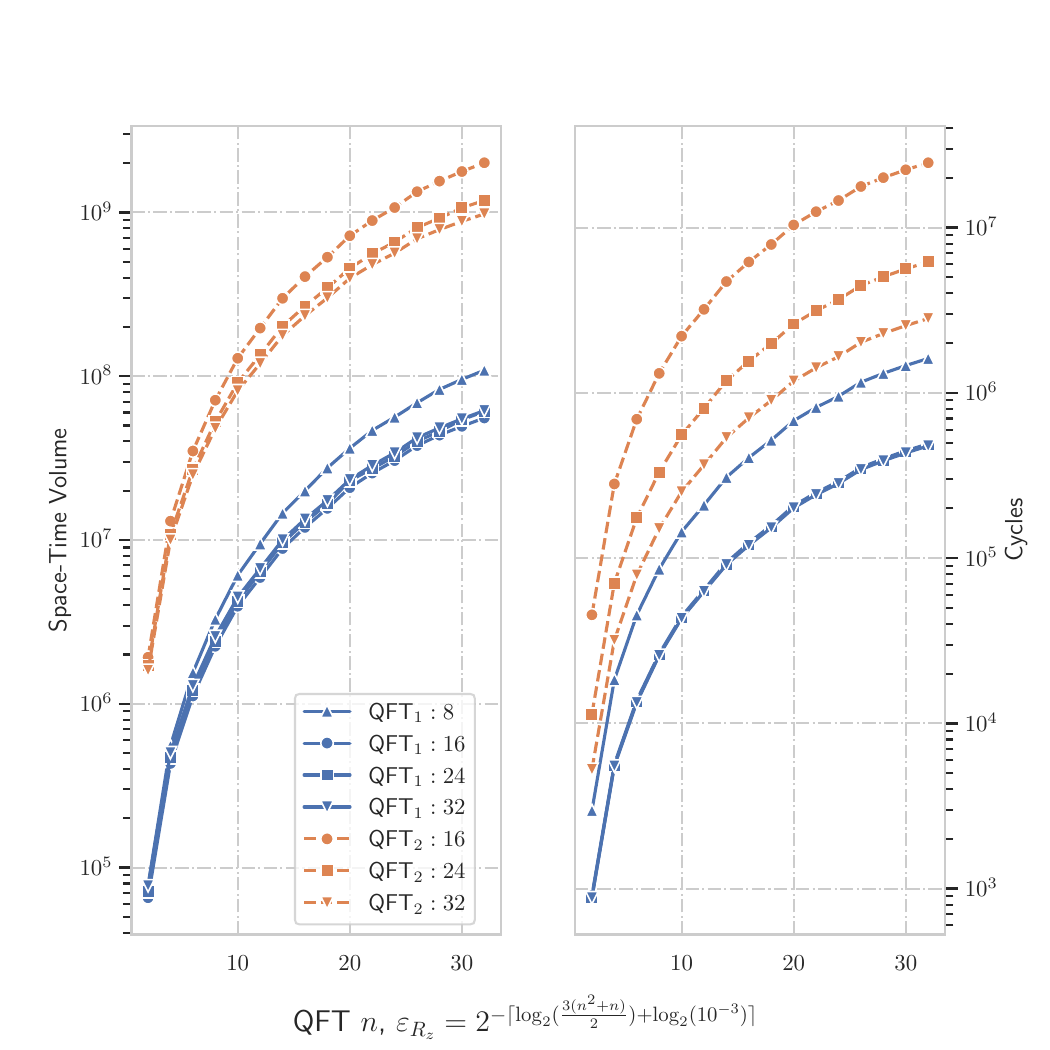}
\end{center}
    \caption{Space-Time Volume and Cycle costs for a banked-cell QFT implementation where the legend depicts the number of distillation rounds for the $T$ gates along with the size of the $CR_Z$ extern. \label{fig:mcx_results}
}
\end{figure}

The results of this compilation can be seen in \Cref{fig:mcx_results}.
Mirroring the results of the Toffoli network benchmark, the reduced space-time volume and runtime of the CCZ factories compared to Toffoli gates decomposed to $T$ gates is the primary differentiator.  
Also as before these results obfusate the relative fidelities of the $T$ and $CCZ$ states. 

Otherwise unrolled DAGs outperform their equivalent externs, at the cost of greater compilation complexity.

\subsubsection{Quantum Fourier Transform}

The QFT is constructed from a collection of $CR_Z(\theta)$ rotations. 
Each $CR_Z(\theta)$ can be constructed using two $R_Z(\theta / 2)$ operations, a $R_Z(-\theta / 2)$ along with a pair of CNOTs. 

This is ripe for compilation with an extern, and for setting the mapping heuristic to match on extern sizing. 
By compiling all externs to the same unital size we are left with compiling a fixed number of $CR_Z$ externs.    

Within the $CR_Z$ extern we may compile each of the $R_Z$ gates as factories by teleporting the $R_Z$ operation. 

By fixing the bank size to $n$ externs for an $n$ qubit QFT, we compare the performance of the QFT as a function of the size of the $CR_Z$ externs.     
 
For this QFT we construct a bank of dedicated $R_Z(\theta)$ externs of equal size for the range of thetas required, with the precision set such that $\sum_i \varepsilon_{RZ(\theta_i)} < 10^-3$. 
These externs are dynamically allocated at runtime so that any $CR_Z(\theta)$ operation may be assigned to any extern, leveraging that each $CR_Z$ operation has been compiled to an extern of the same size. 
Unlike in the previous examples where the total QCB size varies, here we vary the size of the $CR_Z(\theta)$ QCB elements.

\subsection{Arithmetic Operations}
\label{sec:evaluation/arithmetic}

Quantum arithmetic circuits underpin a collection of quantum algorithms.
 Shor's algorithm assumes a quantum modulo power operation, while a number of quantum simulation algorithms depend on the implementation of Taylor series calculation on quantum hardware.

To construct these arithmetic circuits we have built a quantum multi-precision arithmetic library.

The library performs classical simulation of CNOT, Toffoli and $X$ gates on binary vectors, and implements a linear allocator for ancillae tracking and re-use.
 The classical verification of these operations is comparatively computationally cheap.
 Once verified each of the gates of these arithmetic circuits has then been mapped to a sequence of surface code operations and passes to the compiler.

This section details the costing of a range of quantum circuits for varying QCB sizes.

\subsubsection{$n$-bit Addition and Subtraction}

Ripple addition and subtraction circuits are inverses of each other, and each are constructed with a sequence of `MAJ' and `UMAJ' gates~\cite{cuccaro_adder}.

These circuits may implement either modulo arithmetic or increase the size of the output register by a qubit to track an overflow bit. An example circuit may be seen in \Cref{fig:adder_example}.

As this implementation of an adder is not particularly parallelisable, the rate of magic state consumption by the circuit is approximately constant, and depends on the Toffoli gates.
Additionally as the UMA and UMAJ sequence lengths increase with the size of the register, but the rate of $T$ gates remains constant a constant number of magic state factories are required to implement this circuit within optimal time bounds for any size of adder or runtime of factory.

\begin{figure}
\begin{center}
    \includegraphics[scale=0.5]{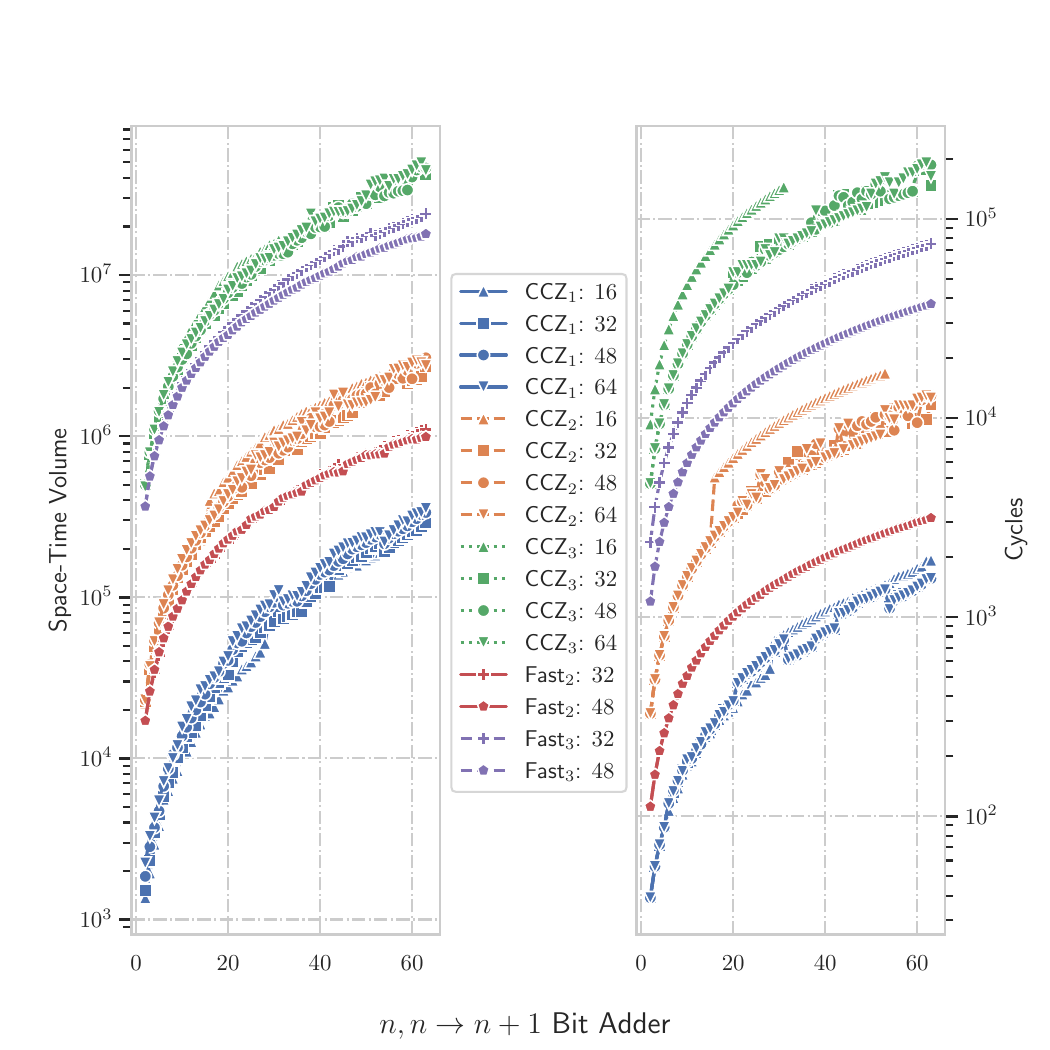}
\end{center}
    \caption{Cycle cost of addition circuits with one-level $15$-$1$ factories. The behaviour of the $12 \times 12$ and the $16 \times 16$ QCBs is caused by the increased memory requirements of the larger addition circuits forcing the compiler to allocate one fewer factory. The upper bound on the runtime then represents the single factory case. A more extreme version can be seen for the $10 \times 10$ compilation where for a 14 qubit or larger adder the compiler rejects the input on the grounds that there is insufficient memory for registers and routing.   
\label{fig:results_adder}}
\end{figure}

Unfortunately, for a fixed size QCB, as the register size increases the memory required for registers, and hence available for dedicating to externs decreases.
Eventually this results in a sub-optimal number of magic state factories and reciprocally impacts the runtime of the adder.
This can be seen in \Cref{fig:results_adder}.
As the size of the adder increases the $16 \times 16$ QCB eventually reports an allocation error due to insufficient size. 
As the factory size increases this error is reported for smaller and smaller registers.

Lastly we compile two additional sets of CCZ$_n$ factories, with four CCZ$_{n - 1}$ allocated externs. These are denoted as `Fast$_i$'.  
For the adders, the additional space taken by these larger factories is offset by reductions in idling time that outweigh the reduction in parallelism.    

Whereas the Toffoli gates exhibited only a mild improvement in space time volume with larger QCB sizes, the adders contend with the space time volume of the additional $3n$ idling qubits.  
This leads to a range of algorithm-specific extern optimisation, where the size of nested externs impacts the performance of the circuit.       

\subsubsection{Multiplication}

A shift and add multiplication circuit is an out of place operation that takes two registers of size $a$ and $b$ and writes the output to a register of size $a + b + 1$.
This is performed by the sequential application of conditional addition circuits with shifted outputs.
The conditional additions are performed by using Toffoli gates as a set of `controlled copy' operations, such that a register may be conditionally copied, before addition is performed\cite{Vedral_1996}.
By manipulating the scope of the target register the output of the addition is raised by a integer power of two.

If the output register is guaranteed to start in the $\ket{0}$ state, then each round of addition uses an adder of size $a$ or $b$.
If this condition is not met then the round adder for a round $n$ should be of size $a + b + 1 - n$.
This requires additional ancillae qubits for the intermediary copy register.

\begin{figure}
\begin{center}
    \includegraphics[scale=0.5]{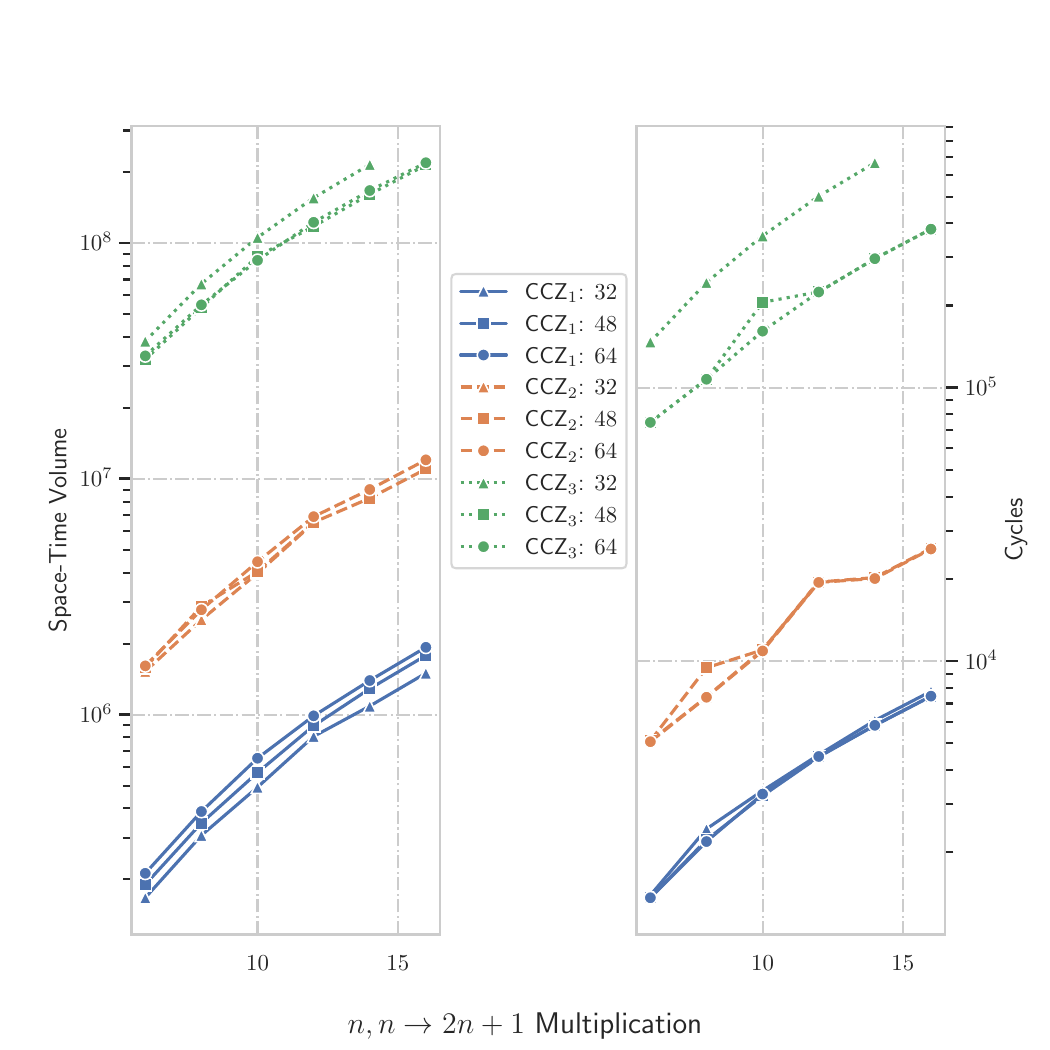}
\end{center}
    \caption{Space-Time Volume and Cycle costs for a $n$ Qubit Multiplication using $T^{8} \rightarrow CCZ$ factories. \label{fig:mul_ccz}}
\end{figure}

As with the addition circuit, most of the shift-and-add multiplier is highly linear. 

We have constructed our multiplication by first composing a controlled copy extern, along with an adder of the appropriate size.      
Multiplication then involves invoking instances of these two externs. 
As each extern shares dependencies, the programme is highly sequential, limiting the
The main advantage of allocating multiple Adder or CCPY externs is to reduce movement times for registers between these objects, otherwise the costing for this circuit with the given construction is mostly a bound on the number of allocatable registers.

\subsubsection{Integer Division}

Our division circuit implements integer division via trial division.
. Our inputs to this circuit are a dividend of length $a$, a divisor of length $b$, and our outputs are the two inputs, along with a quotient of length $a - b + 1$ and a remainder of length $a + 1$. This circuit is similar to the implementation in~\cite{thapliyal2018quantum}.

Division begins by copying the top $a - b$ qubits from the dividend register to the remainder. Each round of the division circuit then compares whether the dividend or the remainder is larger, by subtracting one from the other. If the dividend is larger and the high bit is not flipped then the appropriate bit in the quotient is flipped, otherwise the subtraction is reversed. At the start of each round after the first another bit is added from the dividend to the remainder.  

\begin{figure}
\begin{center}
    \includegraphics[scale=0.5]{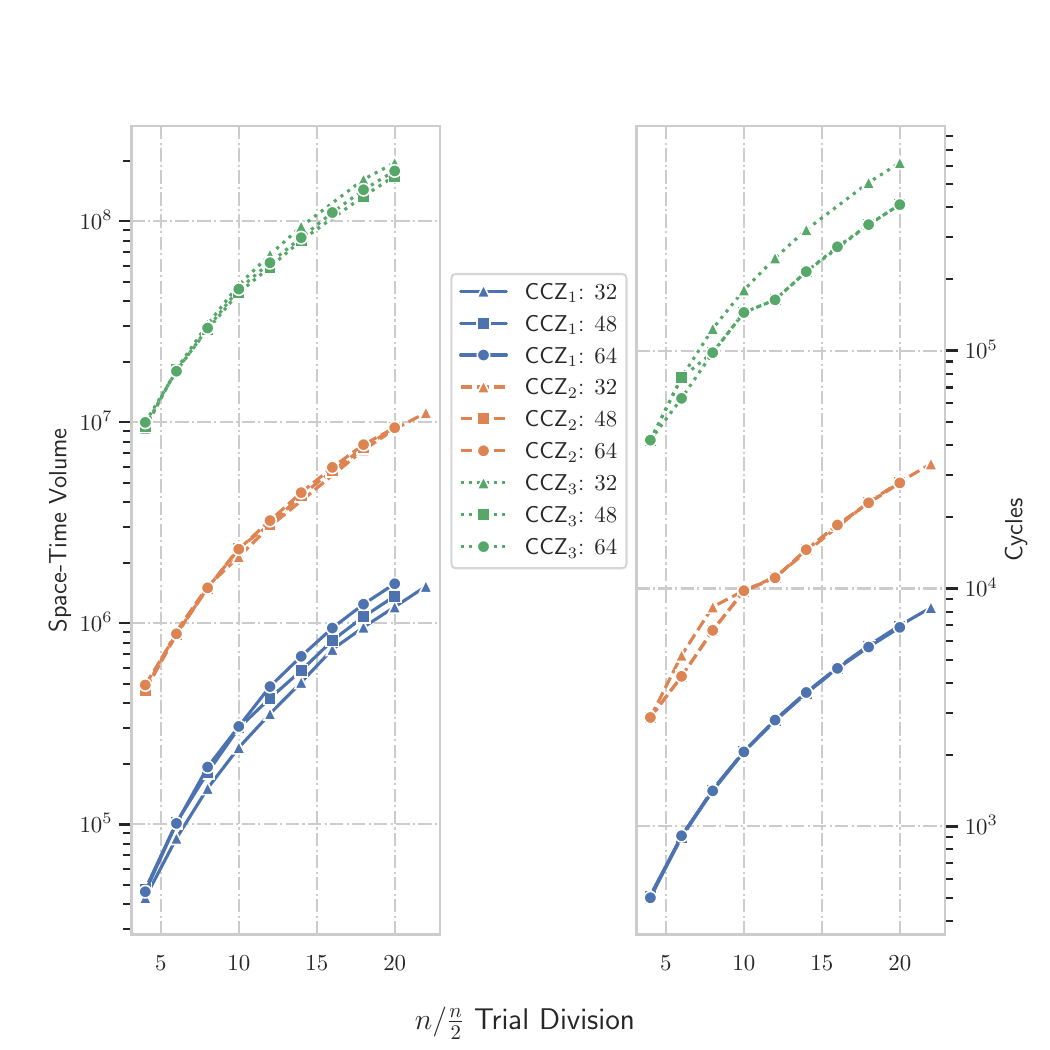}
\end{center}
    \caption{Performance of $n$ qubit division circuits for a relative divisor register size of $n / 2$. \label{fig:qmpa_div} }
\end{figure}

The performance of the division circuit can be seen in \Cref{fig:qmpa_div}. As the division circuit depends on repeated applications of the addition circuit, the same pattern of reduced performance is observed when the rate at which magic states are required exceeds the rate at which they are produced for a given QCB size.    

\subsection{Memory Operations}
\label{sec:evaluation/memory}

Another useful set of operation to benchmark are quantum memories. 
Here we will consider two different QRAM implementations\cite{qram_critique}.

\subsubsection{Bucket Brigade}

\begin{figure}
\begin{center}
    \begin{subfigure}{0.22\textwidth}
        \begin{centering}
    \includegraphics[scale=1]{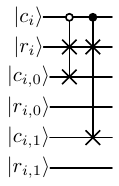}
        \end{centering}
    \end{subfigure}
    \begin{subfigure}{0.22\textwidth}
        \begin{centering}
    \includegraphics[scale=1]{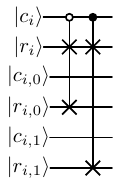}
        \end{centering}
    \end{subfigure}
\end{center}
    \caption{Bucket Brigade QRAM control and route gadgets. These circuits are identical up to a permutation of the input qubits. They are compiled to the same extern object, the performance of which can be seen in \Cref{fig:bb_results}. A Bucket Brigade QRAM is constructed from a sequence of these operations.\cite{qram} \label{fig:qram-bb}}
\end{figure}

\begin{figure}
\begin{center}
    \includegraphics[scale=0.45]{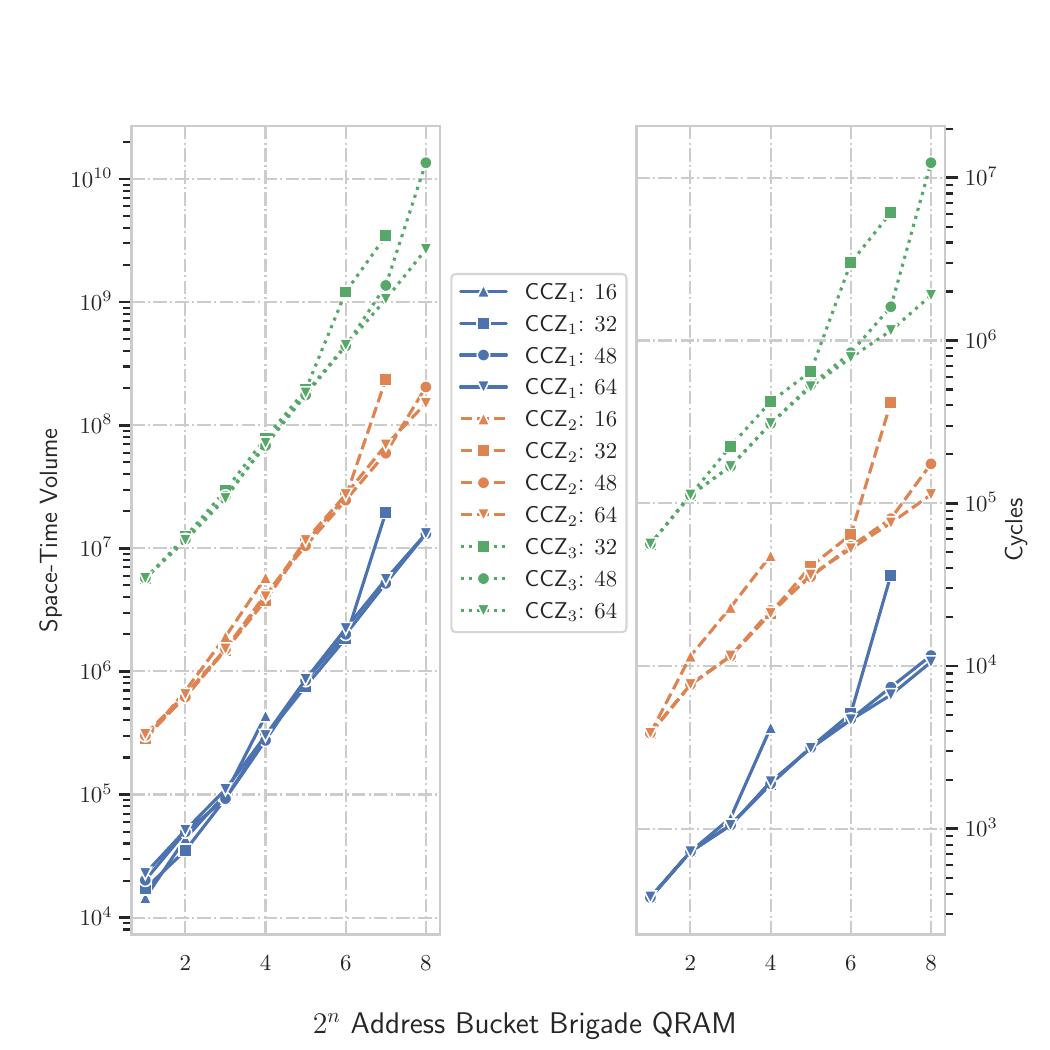}
\end{center}
    \caption{Bucket Brigade QRAM Gadget and Readout performance. The bucket brigade gadget is the set of controlled swaps in \Cref{fig:qram-bb}\label{fig:bb_results}. The forward routing and readout circuit can be seen in \Cref{fig:bb_circuit}}
\end{figure}

The bucket brigade QRAM depends on a series of pairs of controlled swaps as seen in \Cref{fig:qram-bb}\cite{qram, qram_critique, qram_robustness}.
Each of these control and routing circuits are implemented as an extern (the same extern, with different arguments passed to it).
This bucket brigade `gadget' does not scale with the size of the register or the address of the QRAM.
 However, as the controlled swap operations require a sequence of Toffoli gates the runtime of the gadget depends on the number of magic state factories that are available.
This dependency can be seen in \Cref{fig:bb_results}, where the performance of the QCB jumps stochastically when there is enough room to admit another magic state factory.

\begin{figure*}
\begin{center}
    \includegraphics[scale=0.47]{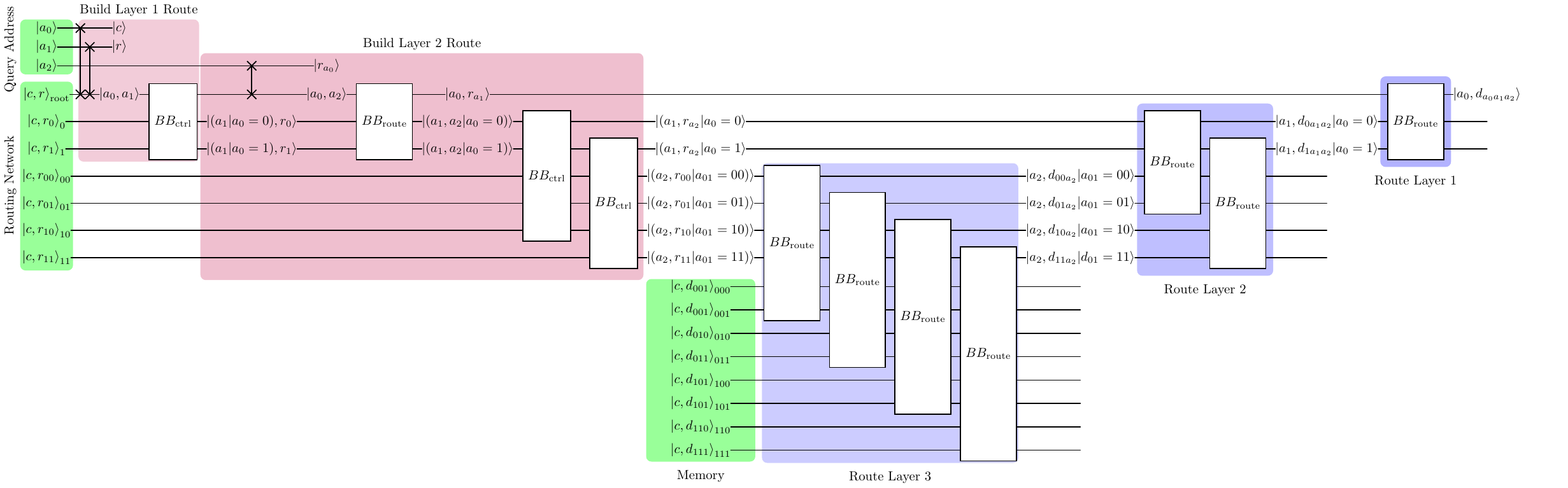}
\end{center}
    \caption{Bucket Brigade QRAM route and readout circuit. The $BB_{\text{route}}$ and $BB_{\text{ctrl}}$ circuits may be seen in \Cref{fig:qram-bb}. Once a readout has been performed the routing network is uncomputed.  \label{fig:bb_circuit}}
\end{figure*}

\begin{figure}
\begin{center}
    \includegraphics[scale=0.5]{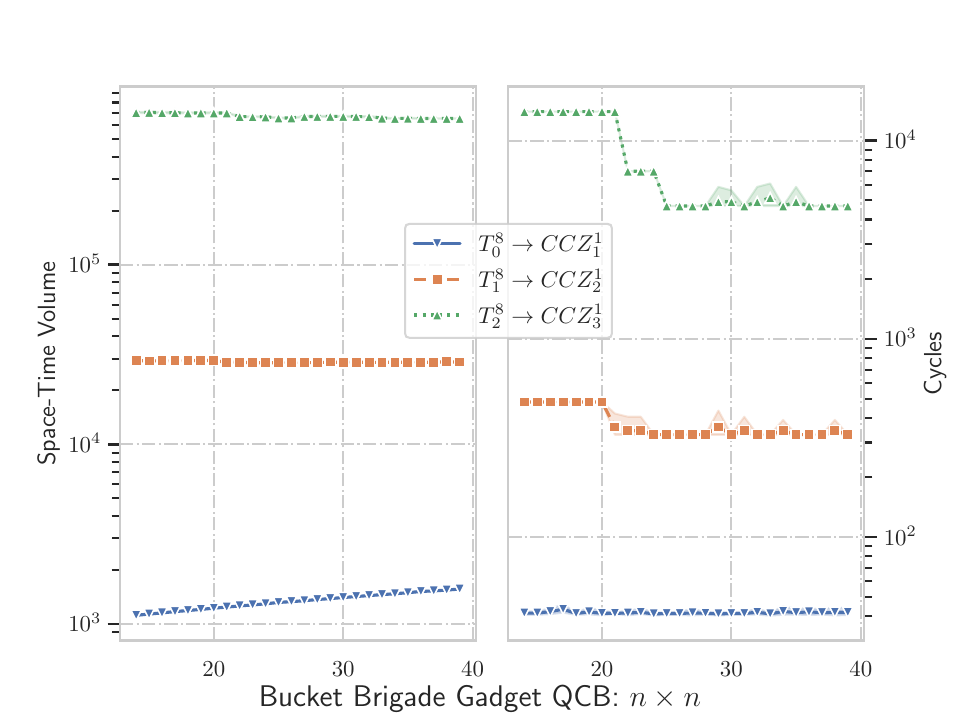}
\end{center}
    \caption{Bucket Brigade QRAM Gadget performance. The bucket brigade gadget is the set of controlled swaps in \Cref{fig:qram-bb}. Each CSWAP operation is decomposed into three Toffoli gates, which in turn depend on a $CCZ$ factory.  \label{fig:bb_gadget_results}}
\end{figure}

The bucket brigade gadgets are benchmarked in \Cref{fig:bb_gadget_results}. 

The performance of the bucket brigade QRAM itself also depends on the number of extern gadgets that are available, and can be seen in \Cref{fig:bb_results}.
The construction of the routing network and the readout are highly parallelisable, as subsequent rounds of the routing network may begin while the current round is still being propagated.
However the dominant memory cost is the exponential number of registers that are required for a $n$ addressable QRAM. 
This set of registers eventually renders a fixed size QCB inoperable as the memory requirements render the allocation of routes and externs impossible.    

The QCB size then determines both the parallelisability of the QRAM and sets the cutoff for a given address size.

In direct contrast to the smaller circuits, the improvement in idling costs from larger QCB for the QRAM translates to marked improvements in space -time volumes.

\subsubsection{Fanout and Swap}

Similarly to the bucket brigade, we construct a fanout and swap gadget, comprised of a set of $k$ controlled swaps\cite{qram_critique, qram_yuan}.
Each gadget takes in two banks of memory of size $k$ registers and a control ancillae, and performs the swap using Toffoli gates.
The fanout and swap QRAM circuit may be seen in \Cref{fig:fanout_circuit}.  
With sufficient gadgets the runtime of the fanout and swap QRAM is $2n$ gadgets. 
The runtime of the circuit is constrained by the rate at which externs are required compared to their availability.

Unlike the bucket brigade, the fanout and swap query is performed in place on the memory block rather than on a separate routing network. This reduced memory footprint admits additional room for externs, which in turn improves the runtime.

\begin{figure}
\begin{center}
\includegraphics[scale=0.45]{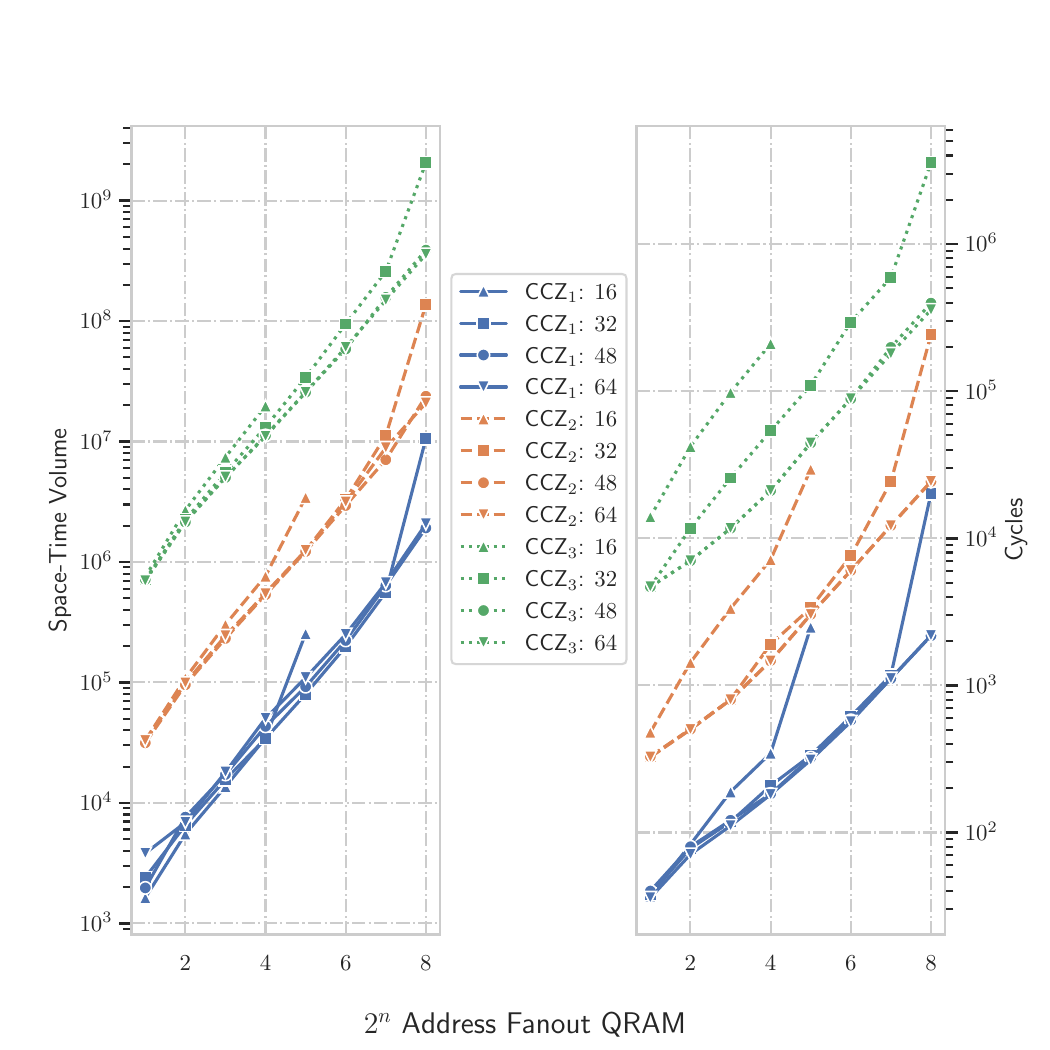}
\end{center}
    \caption{Runtime for fanout and swap QRAM gadgets and readout. Speedups from parallelism are constrained by the size of the QCB, and the number of gadgets that may be allocated. The gadget's size and runtime scales with the size of the registers in the QRAM, and the number of gadgets required scales with the size of the address. \label{fig:fs_results}}
\end{figure}

This interaction may be seen in \Cref{fig:fs_results}.
As with the bucket brigade, the fanout and swap exhibits the same characteristic parallelisability, and exponential register cutoff.

\section{Discussion}
\label{sec:discussion}

The current compiler design has several limitations.
The primary concern is that while the compilation is polynomial in the size of the DAG and the size of the QCB, that for large instance sizes that the implementation becomes slow.   
The primary bottleneck here is the repeated computation of the DAG evaluation heuristic, which is run for each optimisation placement, and requires the re-consumption of the DAG.   
Replacing this heuristic may provide performance improvements that may more readily support larger instance sizes.  

The abstraction of both externs and macros as DAG operations leads to the ability to hot-swap between macro expansion and extern implementations of the same gates, allowing for comparisons without re-writing large amounts of code to account for the change in memory structures within the QCB.
Similarly this allows for swapping between different externs that satisfy the same dependency, without rewriting the DAG.  

A trade-off that has been exhibited for several programmes is between idling and routing overheads - where large QCBs admit improved parallelism, while smaller QCBs have reduced routing overheads.  
As in many cases the dominant cost is a function of the choice of externs, selecting a QCB size to minimise the overall space-time volume is sensitive to both the function being compiled, and the number of rounds of magic state factories - themselves determined by the global problem size.

Another constraint is that the compiler assumes that gates are deterministic, or maybe be accompanied by correction operations that resolve to deterministic gates.   
These assumptions are not upheld by recent advances in $T$ state generation techniques~\cite{lee2025lowoverheadmagicstatedistillation,gidney2024magicstatecultivationgrowing}.

As gates are scheduled in an ASAP order, registers are initialised in their first possible cycle - leading to additional idling overheads that could be resolved by a hybrid ASAP/ALAP scheduling.

Future work would involve further mapping externs to class-like objects, extending the notion of the dynamic allocation and lifetime of the object.
Similarly, the notion of ownership of data would assist in physical register movement between externs, reducing routing costs by improving memory locality.  
The natural extension of this would be compiling to a sequence of externs and abutments, in a facsimile to VLSI.   

Lastly relaxing the QCB correctness requirement that at least one instance of each extern exists, to instead guarantee that a region of sufficient size exists for each extern instance would result in significant improvements in memory-usage and potential improvements in parallelisation. 
The quandary that this situation introduces is determining the size of regions to allocate.

\section{Conclusion}
\label{sec:conclusion}

In this paper we have exhibited a dynamic allocation method for lattice surgery based quantum compilation with fixed memory constraints.  
This represents an advancement over hand-designed regions\cite{opensurgery}, or over architecture-unaware benchmarking methodologies\cite{AQREpaper,AQRE}.    
For large, utility scale quantum algorithms the need for automatic compilation routines arises from the infeasibility of hand crafting layouts for large ranges of different applications. 
 Optimising resource allocations within fixed memory constraints abstracts away explicit resource concerns in terms of the number of instances of factories, and instead draw on hardware resource constraints, within the abstractions provided by the surface code.

\section*{Acknowledgements}
This work is supported by the Defence Advanced Research Projects Agency (DARPA)
Quantum Benchmark program under grant numbers HR00112230007, HR001121S0026, and HR001122C0074.
Any opinions, findings and conclusions or recommendations expressed in this material
are those of the authors and do not necessarily reflect the views of DARPA.
YRS acknowledges the unhelpful contributions of Shimon Arjun Kiefer-Sanders,
whose birth considerably delayed the release of this manuscript.

\bibliographystyle{abbrv}
\bibliography{references}


\appendix

\section{QCB Allocation}
\label{sec:register_placement}


\subsection{Register Placement}

Register placement follows the process described in \Cref{tbl:register_placement_rules}.
Extern placement follows the process described in \Cref{tbl:extern_placement_rules}.

The coordination of these placement rules is discussed in \Cref{apx:initial_placement} and \ref{apx:opt_placement}.

\begin{table*}
        \centering
        \begin{tabular}{cp{55mm}p{55mm}M{15mm}}
           \toprule
  \textbf{Rule} & \textbf{Condition} & \textbf{Instruction} & \textbf{Example} \\ \midrule
            \midrule
             1 & Placement is in top left corner of QCB. & Place an register and place a routing row directly beneath it.  & \includegraphics[width=1.75cm]{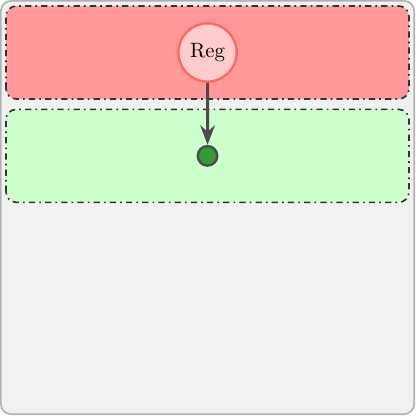} \\\midrule
             2 & Placement is on the top of the QCB & Drop a route from the left. & \includegraphics[width=1.75cm]{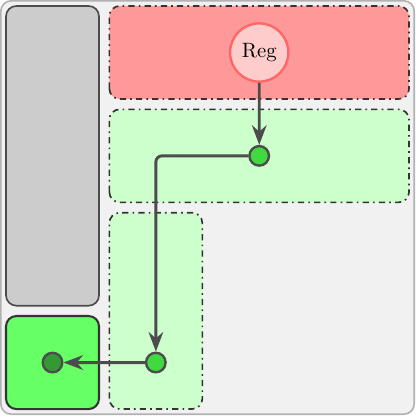}\\\midrule

             3 & Bus above placement. & Place register and leave route on left side of placement.  & \includegraphics[width=1.75cm]{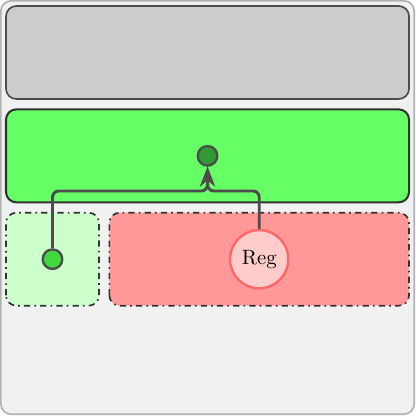}\\\midrule

             4 & Top left of region is connected to the bus. & Place a register with a route to the left of and below the register. &  \includegraphics[width=1.75cm]{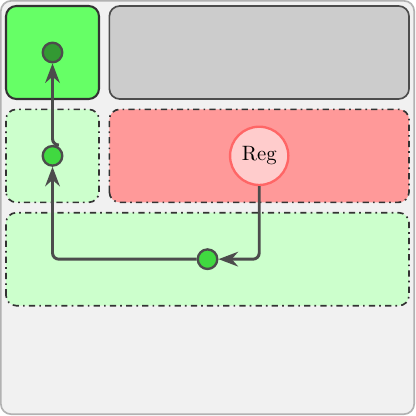}\\\midrule


 5 & & Attempt to route up the left hand side of the register and connect with the bus. & \includegraphics[width=1.75cm]{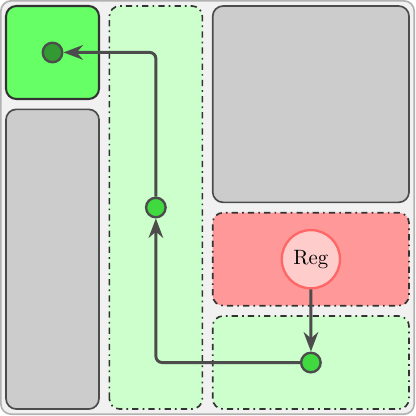}\\\midrule
 6 &  & Attempt to route down the left hand side of the register and connect with the bus. & \includegraphics[width=1.75cm]{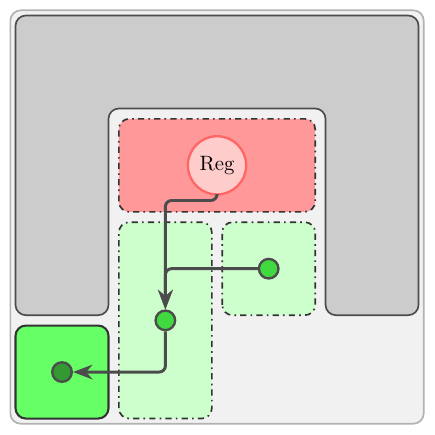}\\\midrule
            \bottomrule
        \end{tabular}
        \caption{Register Placement Rules. Each placement is tried in order, the first legal placement is then accepted.}
        \label{tbl:register_placement_rules}
\end{table*}

\subsubsection{ IO placement}
If the QCB requires any IO elements they are placed after the first register or route. 

\begin{enumerate}
    \item Place an IO segment with one register for each IO element from the bottom left corner of the QCB.
    \item If possible place a route segment above the IO, if this is not possible then assert that each surface code patch above the IO segment is a routing patch.      
    \item If possible place a route segment to the right of the IO and join it to the route above the IO
\end{enumerate}

These operations ensure that the IO will eventually be joined to the routing network.
Until the join occurs, the routing elements associated with the IO segment are not considered to be part of the routing bus.

By considering the final placement on the left edge of the QCB and directly above the IO route the set of legal IO connections can be demonstrated.   

\begin{table*}
        \centering
        \begin{tabular}{cp{55mm}p{55mm}M{15mm}}
           \toprule
  \textbf{Rule} & \textbf{Condition} & \textbf{Instruction} & \textbf{Example} \\ \midrule
            \midrule
             1 & Placement is in top left corner of QCB. & Place an extern and place a routing row directly beneath it.  & \includegraphics[width=1.75cm]{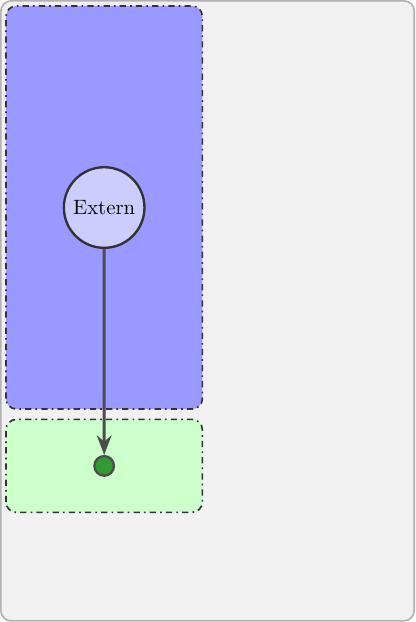} \\\midrule
             2 & Placement is already routed from below. & Add the extern segment above the route. & \includegraphics[width=1.75cm]{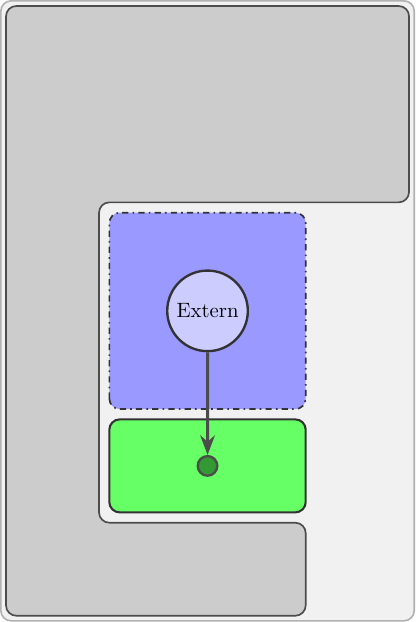} \\\midrule
             3 & Placing a route below would already connect with another routing element. & Add the extern and the route. & \includegraphics[width=1.75cm]{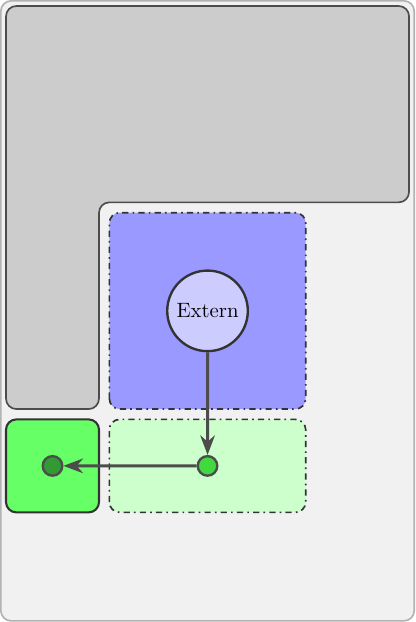}\\\midrule
             4 & Placement is on the top row of the QCB. & Attempt to place extern and routing row. Attempt to connect route downwards from the left to the prior allocation.   & \includegraphics[width=1.75cm]{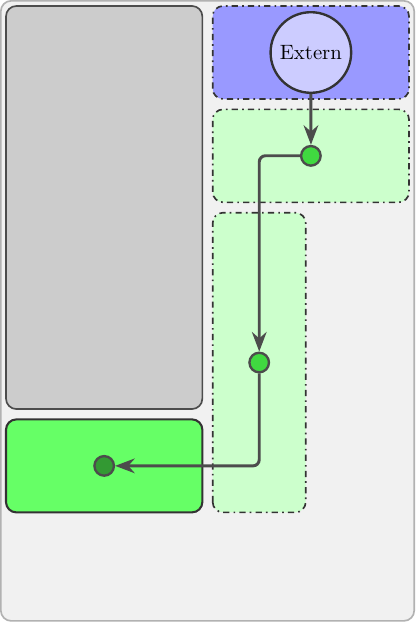}\\\midrule
             5 & Placement is on the top row of the QCB. &  Attempt to place extern and routing row. Attempt to connect route from the upwards from  left to the prior allocation.& \includegraphics[width=1.75cm]{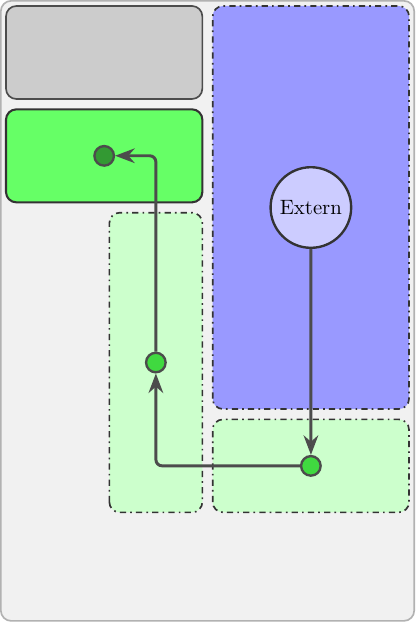}\\\midrule
 6 & & Attempt to route up the right hand side of the extern and connect with any existing route. & \includegraphics[width=1.75cm]{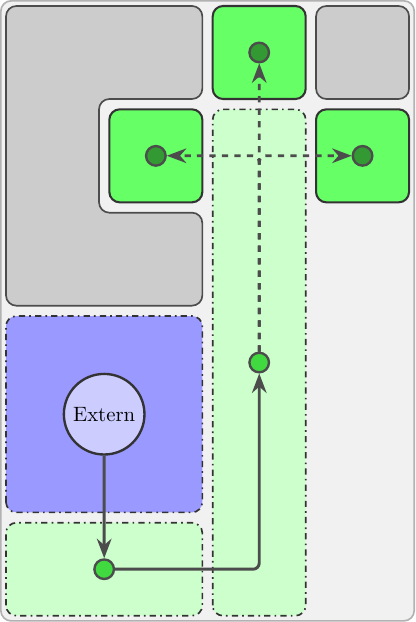}\\\midrule
 7 &  & Attempt to route up the left hand side of the extern and connect with any existing route. & \includegraphics[width=1.75cm]{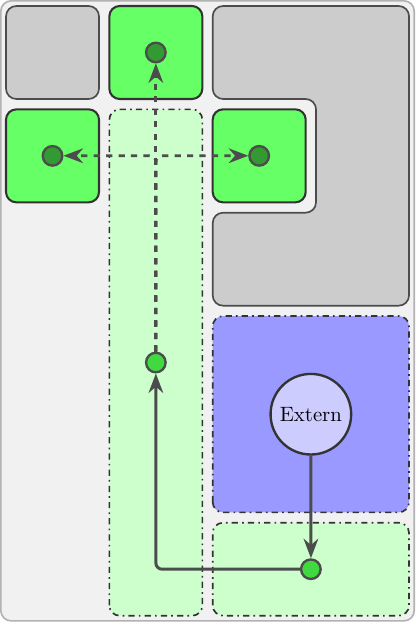}\\\midrule
8 &  & Attempt to route down from the left hand side of the extern and connect with any existing route. & \includegraphics[width=1.75cm]{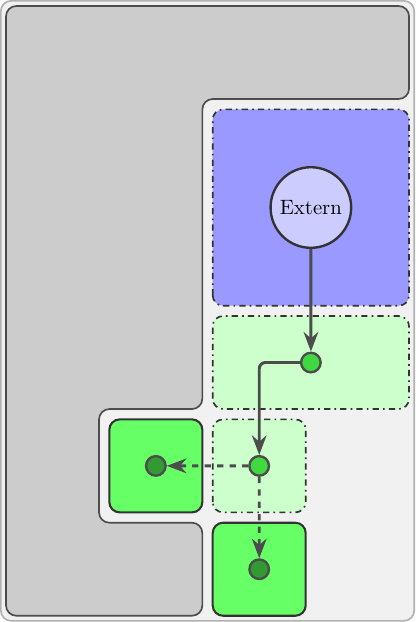}\\\midrule
            \bottomrule
        \end{tabular}
        \caption{Extern Placement Rules. Each placement is tried in order, the first legal placement is then accepted.}
        \label{tbl:extern_placement_rules}
\end{table*}

\subsubsection{Initial Placement Round}
\label{apx:initial_placement}

The initial placement of patches seeks to provide a minimal working implementation of the circuit.
This requires the ability to allocate any externs required for the circuit, a sufficient number of IO elements to support the IO channel of the QCB, a sufficient number of registers patches allocated for each data qubit in the circuit, and the ability to route between all registers, externs and IO channels. 

If a patch placement fails during the initial allocation then compilation reports a failure and halts.

\begin{algorithm}
\caption{Initial QCB Patch Placement}
\begin{algorithmic}
    \IF{Externs is not empty} 
        \STATE Sort Externs by width then height from largest to smallest 
            \STATE Place the first extern segment  
            \STATE Flag placed segment type as allocated 
    \ENDIF
    \IF{IO Patches Required $> 0$}
        \STATE IO Placement 
    \ENDIF
    \WHILE{Not all segment types allocated}
        \STATE Place and flag next unflagged extern
    \ENDWHILE
    \WHILE{Registers Placed $<$ Register Patches Required} 
        \STATE Place a register segment   
        \STATE Registers Placed += Length of placed segment   
    \ENDWHILE
    \IF{IO Patches Required $ > 0$ {\bf and} IO not joined}
        \STATE IO Join 
    \ENDIF
\end{algorithmic}
\end{algorithm}

\begin{algorithm}
\caption{Additional QCB Patch Placement}
\begin{algorithmic}
    \WHILE{Any Placement Succeeds}
        \STATE Attempts = $\text{heuristic}(\text{add bus}, \text{add extern}_i)$     
        \STATE Sort Attempts   
        \WHILE{No Placement Succeeded}
            \IF{Attempting to Place an Extern}
                \STATE Place Extern
            \ELSIF{Attemping to Place a Bus}
                \STATE Split largest Register vertically with route patches   
                \WHILE{Registers Placed $<$ Register Patches Required}
                    \STATE Place Register
                    \IF{Register Placement Fails}  
                        \STATE Bus Placement Fails
                    \ENDIF
                \ENDWHILE
            \ENDIF
        \ENDWHILE
    \ENDWHILE
    \STATE Repeat bus placement until placement fails 
    \STATE Repeat register placement until placement fails  
    \STATE Mark remaining patches as local routes 
\end{algorithmic}
\end{algorithm}

If any of the placement operations fails then the allocation fails and compilation halts. Otherwise if the program completes then the QCB should contain a sufficient number of registers, IO and extern segments to implement the circuit, and the routing elements should support any operation between two allocated addressable patches on the QCB.       

\subsubsection*{Optimisation Placement}
\label{apx:opt_placement}

Having completed the initial placement, we now concern ourselves with the allocation of additional routing lanes and extern segments to speed up computation.

Using our heuristic DAG depth we can speculate on whether allocating further extern segments, or attempting to add additional routing busses will provide a greater speedup.   

Using the DAG heuristic evaluation we sort the set of potential placements by whichever would minimise the circuit depth.  
These placements are then attempted in order until a QCB placement succeeds. 
At this point the process continues until all placements fail, or the heuristic reports no speedups from further placements. 
If there is any remaining space we attempt to place further register segments to exhaust it.
If after the register placement there is still remaining space

\section{Lattice Surgery Operations}
\label{sec:ls_ops}
In this section we sketch out our assumed lattice surgery operation primitives.

\begin{figure}[h]
    \begin{centering}
    \center
    \includegraphics[scale=0.9]{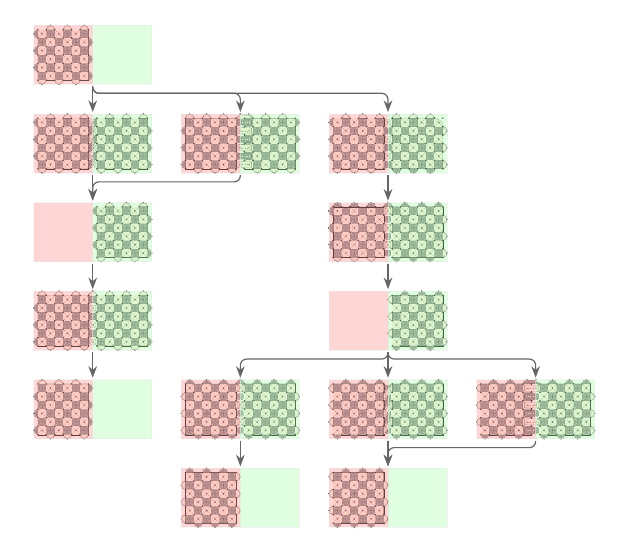}
    \end{centering}
    \caption{Rotation operations for the surface code using corner movement~\cite{gameofsurfacecodes}, twisted~\cite{latticesurgerywithatwist} and stretched~\cite{fowler2019low} stabilisers. All possible single patch rotations can be performed in 3 tocs.\label{fig:rotate_boundary}} 
\end{figure}

\begin{figure}
    \begin{centering}
        \begin{subfigure}{0.22\textwidth}
            \begin{center}
\includegraphics[scale=1.75]{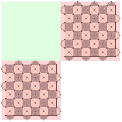}
                \caption{\clock$_0$ Initial State}
                \end{center}
    \end{subfigure}
        \begin{subfigure}{0.22\textwidth}
            \begin{center}
\includegraphics[scale=1.75]{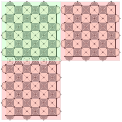}
                \caption{\clock$_1$ Merge control with ancillae.}
               \end{center}
     \end{subfigure}

        \begin{subfigure}{0.22\textwidth}
            \begin{center}
\includegraphics[scale=1.75]{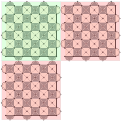}
                \caption{\clock$_1$ Split control and ancillae}
              \end{center}
      \end{subfigure}
        \begin{subfigure}{0.22\textwidth}
            \begin{center}
\includegraphics[scale=1.75]{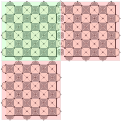}
                \caption{\clock$_2$ Merge target and ancillae}
             \end{center}
       \end{subfigure}

        \begin{subfigure}{0.45\textwidth}
            \begin{center}
\includegraphics[scale=1.75]{img/cnot/surface_code_cnot_0}
                \caption{\clock$_2$ Split and measure out Ancillae}
            \end{center}
        \end{subfigure} 
    \end{centering}
    \caption{Implementation of a CNOT operation on the surface code. The ancillae is prepared in the $\ket{+}$ state, and merge operations are used to perform joint $ZZ$ and $XX$ measurements. These operations implement the circuit in \Cref{fig:cnot_circ}.  } \label{fig:sc_cnot}
\end{figure}

\begin{figure}
    \begin{center}
\includegraphics[scale=0.7]{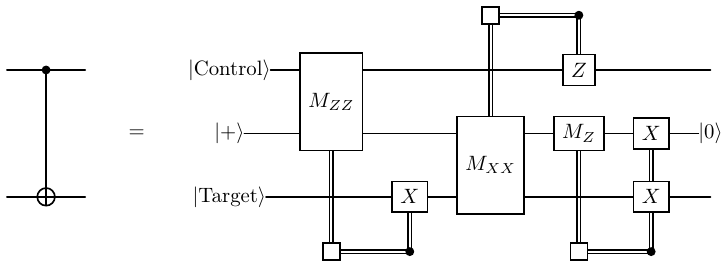}
\includegraphics[scale=0.7]{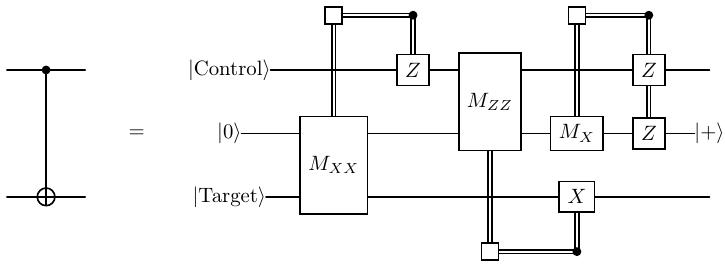}
    \end{center}
    \caption{ Two different implementations of CNOT operations using an ancillae qubit~\cite{lattice_surgery, Lao_2018}.  The joint $Z$ measurement and the joint $X$ measurement are equivalent to rough and smooth merges and splits between surface code patches. \label{fig:cnot_circ}}
\end{figure}

\begin{figure}
    \begin{center}
\includegraphics[scale=0.6]{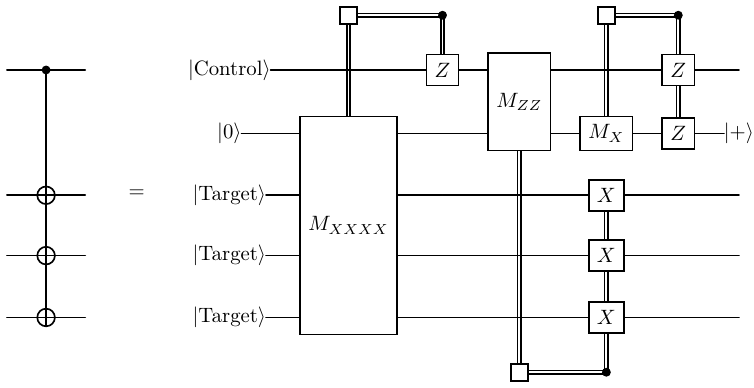}
    \end{center}
    \caption{ Multiple target CX operation using an ancillae qubit. The joint $Z$ measurement and the joint $X$ measurement are equivalent to rough and smooth merges and splits between surface code patches.\cite{fowler2019low}\label{fig:cnot_circ_multi}}
\end{figure}

\begin{figure}
    \begin{center}
\includegraphics[scale=0.7]{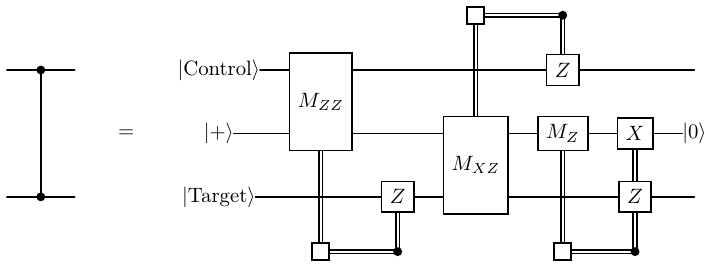}
\includegraphics[scale=0.7]{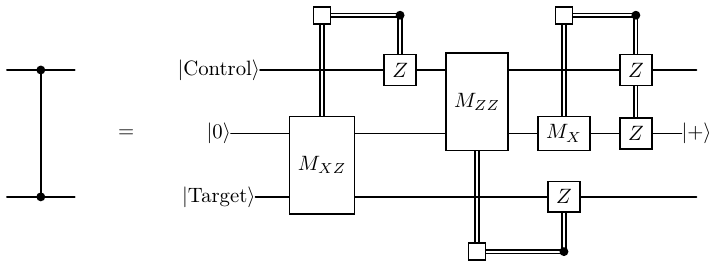}
    \end{center}
    \caption{ Two different implementations of CNOT operations using an ancillae qubit~\cite{lattice_surgery, Lao_2018}.  The joint $Z$ measurement and the joint $X$ measurement are equivalent to rough and smooth merges and splits between surface code patches. \label{fig:cnot_circ}}
\end{figure}

\begin{figure}
    \begin{center}
\includegraphics[scale=0.6]{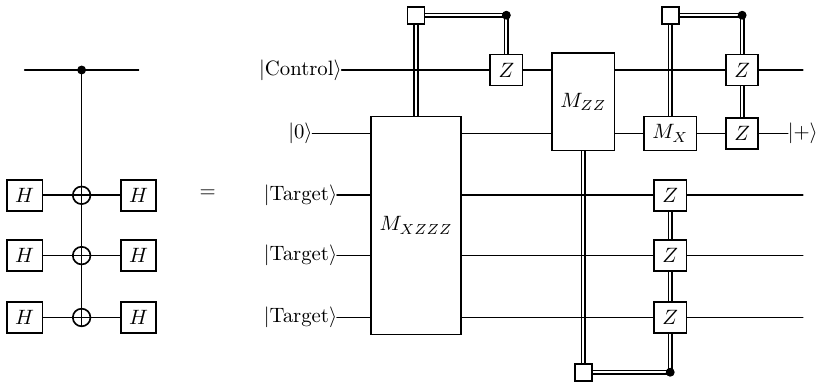}
    \end{center}
    \caption{A multi-target CZ operation\cite{fowler2019low}. \label{fig:cz_circ_multi}}
\end{figure}

\begin{figure*}
    \begin{centering}
        \includegraphics[scale=0.65]{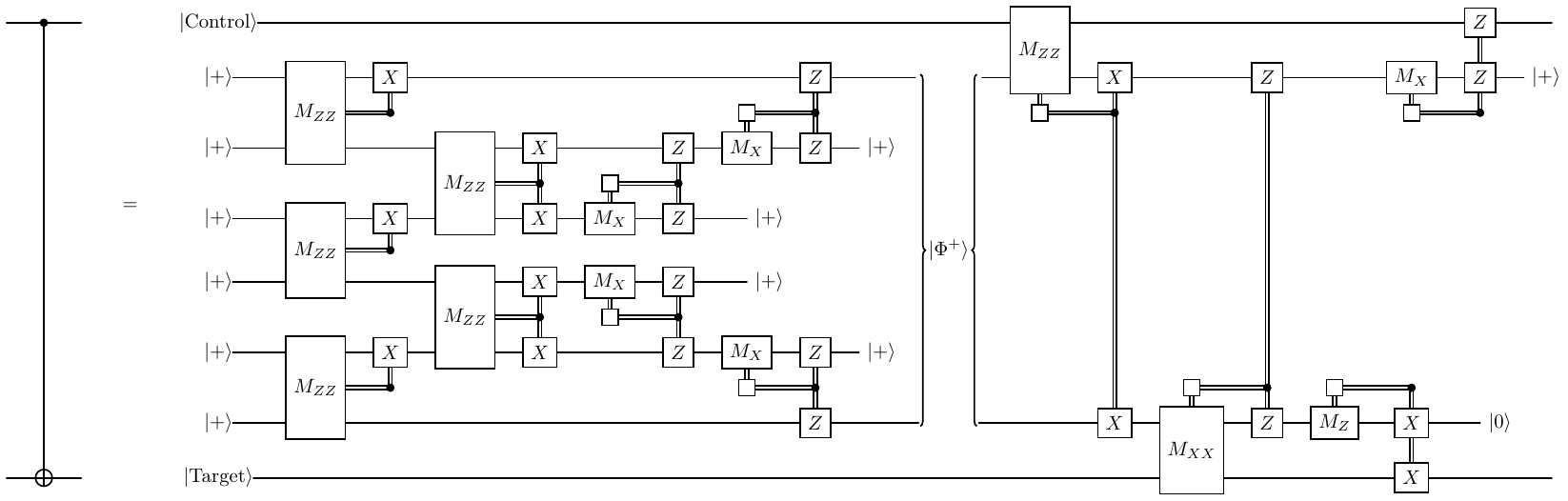}
    \end{centering}
    \caption{Implementation of a non-local CNOT operation using only nearest-neighbour operations along with classical non-local corrections.
    Routing ancillae form a Bell state using the edge-disjoint path construction\cite{Beverland_2022}.
    The ancillae is prepared in the $\ket{+}$ state, and merge operations are used to perform joint $ZZ$ and $XX$ measurements. These operations implement the circuit in \Cref{fig:cnot_circ}. This structure also forms the basis of distributed, multi-target $CX$ and $CZ$ opertions using routing networks. One ancillae is maintained for each logical register. \cite{Beverland_2022}\label{fig:sc_cnot_circ}} 
\end{figure*}

\begin{figure}
    \begin{centering}
        \begin{subfigure}{0.44\textwidth}
\begin{center}
    \includegraphics[scale=0.9]{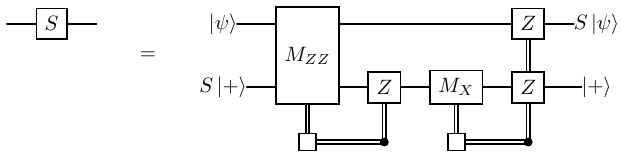}
    \caption{Circuit representation of a phase gate consuming a $\ket{Y} = S\ket{+}$ ancillae state.}
\end{center}
\end{subfigure}
\begin{subfigure}{0.24\textwidth}
\begin{center}
    \includegraphics[scale=1.5]{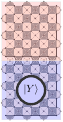}
\end{center}
\end{subfigure}
\begin{subfigure}{0.24\textwidth}
\begin{center}
    \includegraphics[scale=1.5]{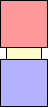} \clock$_1$
\end{center}
\end{subfigure}
    \end{centering}
    \caption{Phase operation where an ancillary $\ket{Y}$ state is consumed~\cite{fowler2013bridge}. 
Prior techniques in fusing twist defects allow for the incorporation of the of the $\ket{Y}$ preparation into the operation~\cite{inplaceY}.
\label{fig:phase_consume}
} 
\end{figure}

\begin{figure}
    \begin{centering}
    \begin{subfigure}[t]{0.45\textwidth}
\begin{center}
        \includegraphics[scale=1]{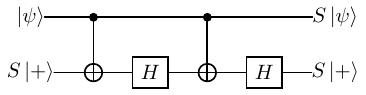}
        \caption{Circuit representation of a Phase gate using a catalysing $\ket{Y} = S\ket{+}$ ancillae state. This implementation preserves the ancillae state without entangling it with the target.} 
\end{center}
    \end{subfigure}

        \begin{subfigure}[t]{0.22\textwidth}
\begin{center}
            \includegraphics[scale=1]{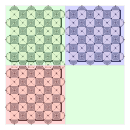}
        \caption{Merge control and ancillae. The state $\ket{\psi}$ is represented in the bottom left corner, while $\ket{Y}$ is in the top right.}
\end{center}
            \end{subfigure}
        \begin{subfigure}[t]{0.22\textwidth}
\begin{center}
    \includegraphics[scale=1]{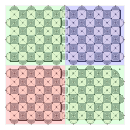}
            \caption{Merge $\ket{Y}$ state and ancillae, completing first CNOT. Start rotation of boundary stabilisers of $\ket{\psi}$}
\end{center}
\end{subfigure}
\begin{subfigure}[t]{0.22\textwidth}
\begin{center}
    \includegraphics[scale=1]{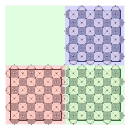}
                \caption{Hadamard $\ket{Y}$ state register. Rotate boundary stabilisers of $\ket{\psi}$}
\end{center}
\end{subfigure}
\begin{subfigure}[t]{0.22\textwidth}
\begin{center}
    \includegraphics[scale=1]{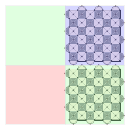}
    \caption{Continue rotation of boundary stabilsiers.}
\end{center}
            \end{subfigure}
\begin{subfigure}[t]{0.22\textwidth}
\begin{center}
    \includegraphics[scale=1]{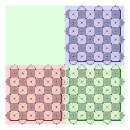}
    \caption{Merge $\ket{\psi}$ with ancillae $\ket{+}$ state.}
\end{center}
            \end{subfigure}
        \begin{subfigure}[t]{0.22\textwidth}
\begin{center}
    \includegraphics[scale=1]{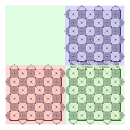}
    \caption{Twisted merge due to offset boundary stabilisers, completing the CNOT operation.}
\end{center}
            \end{subfigure}
    \begin{subfigure}[t]{0.225\textwidth}
\begin{center}
    \includegraphics[scale=1]{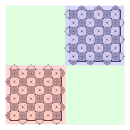}
    \caption{Hadamard on $\ket{Y}$ state.}
\end{center}
            \end{subfigure}
    \begin{subfigure}[t]{0.225\textwidth}
\begin{center}
\includegraphics[scale=1]{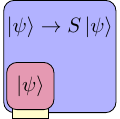} \clock$_7$
\caption{Operation encapsulated as an extern.}
\end{center}
            \end{subfigure}

    \end{centering}
    \caption{Phase operation where an ancillary $\ket{Y}$ state is not consumed and is not entangled with the target state allowing for its re-use. This operation is not used by the compiler, but presents an example of how an extern might be used to encapsulate a hand-compiled operation.}
    \label{fig:phase_duplicate}
\end{figure}

\begin{figure}
\begin{center}
    \includegraphics[scale=0.9]{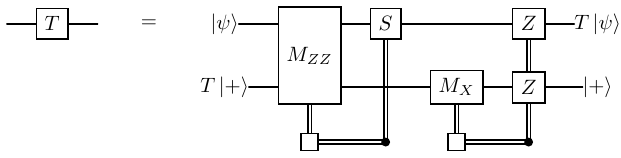}
\end{center}
    \caption{Circuit representation of a T gate consuming a $\ket{T} = T\ket{+}$ ancillae state. The correction operation involving an $S$ gate conditionally requires a $\ket{Y}$ state. }
    \label{fig:t_gate}
\end{figure}


\begin{figure}
\begin{center}
   \begin{tikzpicture}
\draw (0, 0) node[inner sep=0] {\includegraphics[scale=0.35]{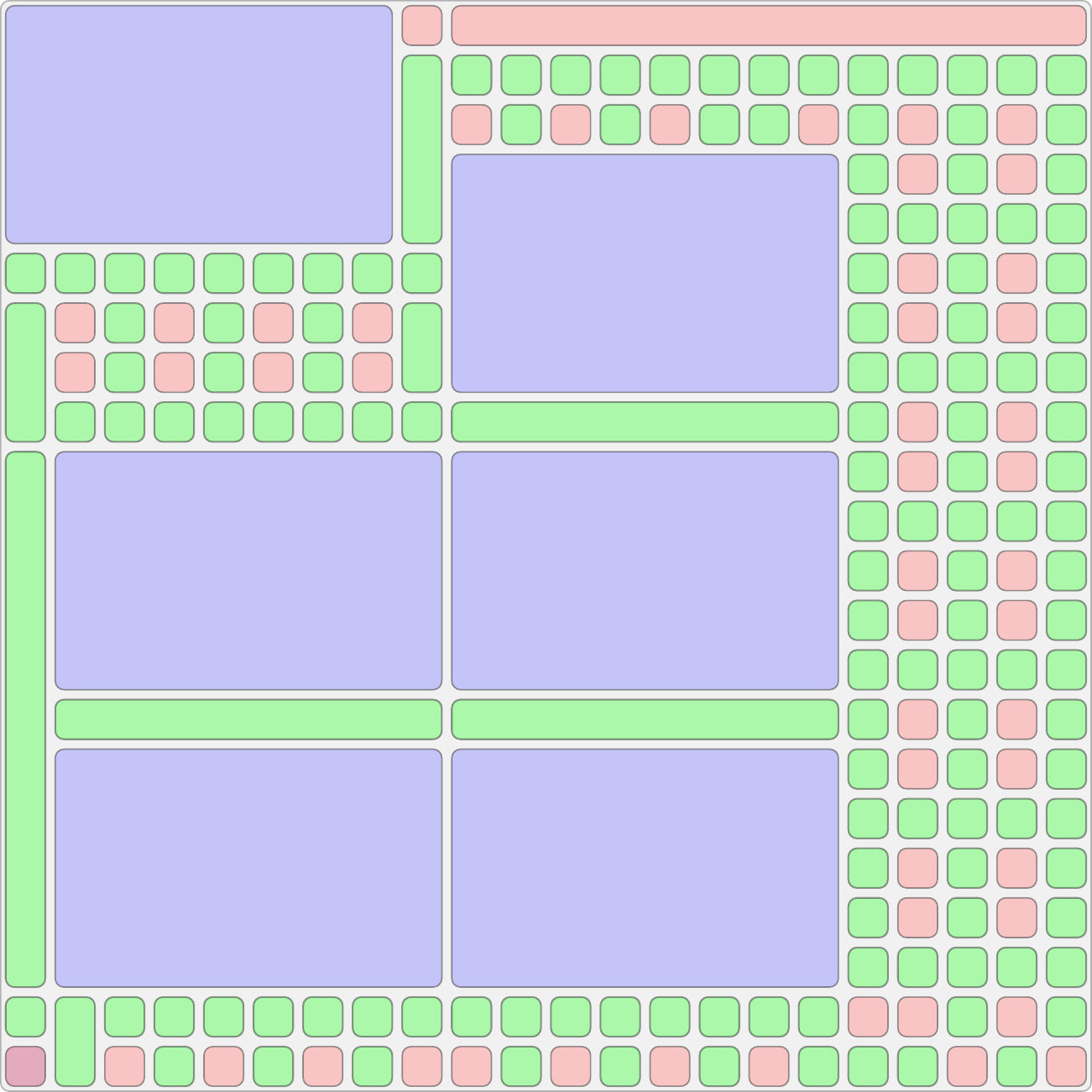}};
\draw (0, 0) node[inner sep=0] {\includegraphics[scale=0.35]{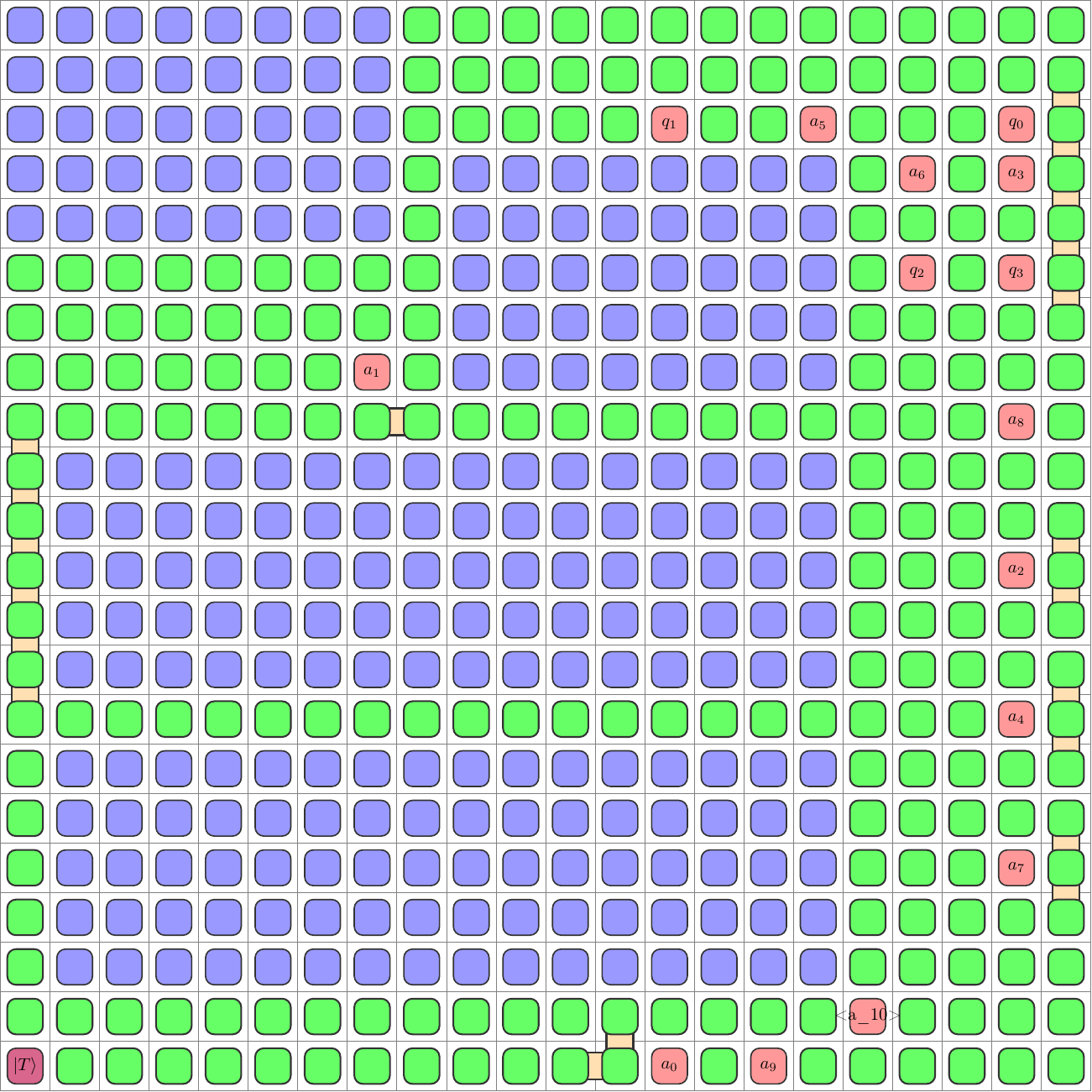}};
   \end{tikzpicture}
\end{center}
    \caption{\small Compiler output showing a QCB implementing a nested $T$ factory containing six externs. An example output from the compilation in \Cref{fig:msf_nest}\label{fig:msf_qcb_example}}
\end{figure}


\begin{figure*}[ht]
\begin{center}
    \includegraphics[scale=0.4]{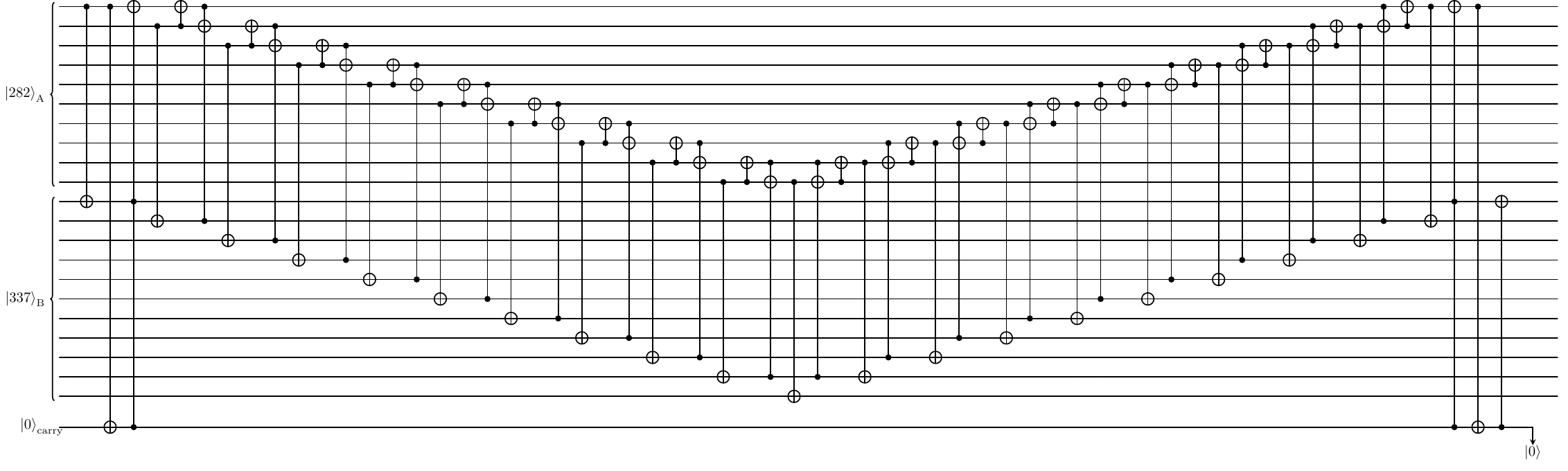}
\end{center}
    \caption{Example Addition Circuit demonstrating the linear sequence of dependent Toffoli and CNOT gates. This version of the adder takes two registers of size $n$ and $n + 1$ qubits, along with an ancillae carry qubit that is reset to the $\ket{0}$ state after the operation of the circuit.\cite{cuccaro_adder} \label{fig:adder_example}}
\end{figure*}

\begin{figure}
\begin{center}
    \begin{subfigure}{0.22\textwidth}
        \begin{centering}
    \includegraphics[scale=1]{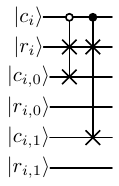}
        \end{centering}
    \end{subfigure}
    \begin{subfigure}{0.22\textwidth}
        \begin{centering}
    \includegraphics[scale=1]{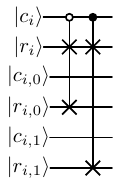}
        \end{centering}
    \end{subfigure}
\end{center}
    \caption{Bucket Brigade QRAM control and route gadgets. These circuits are identical up to a permutation of the input qubits. They are compiled to the same extern object, the performance of which can be seen in \Cref{fig:bb_results}. A Bucket Brigade QRAM is constructed from a sequence of these operations.\cite{qram} \label{fig:qram-bb}}
\end{figure}

\begin{figure*}
\begin{center}
    \includegraphics[scale=0.47]{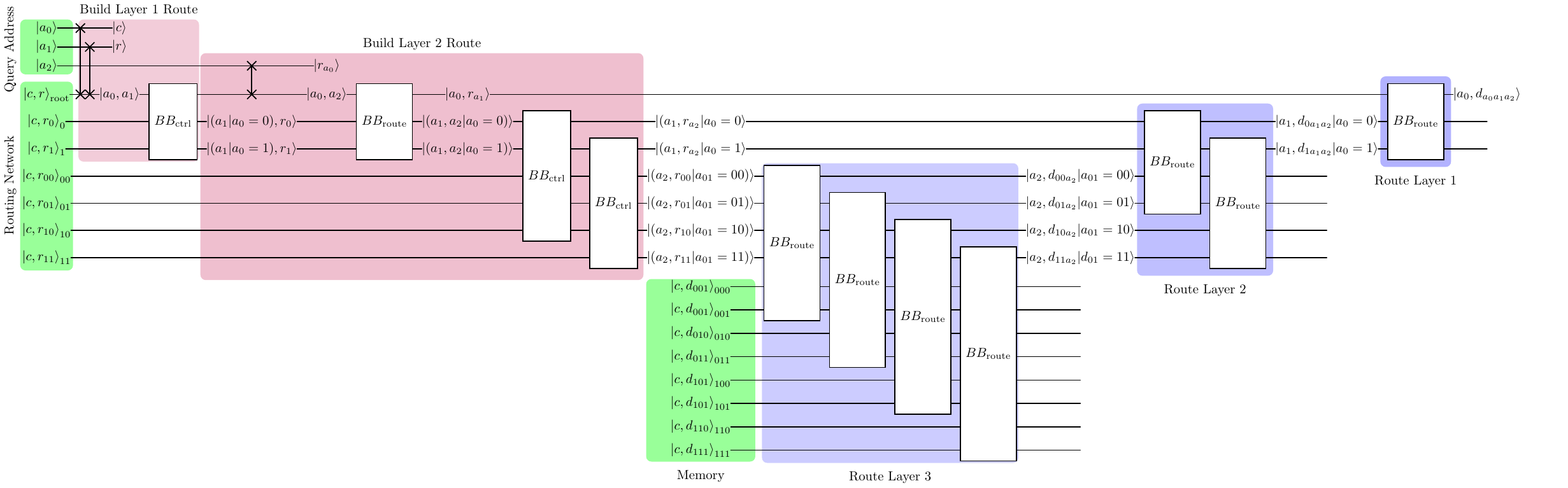}
\end{center}
    \caption{Bucket Brigade QRAM route and readout circuit. The $BB_{\text{route}}$ and $BB_{\text{ctrl}}$ circuits may be seen in \Cref{fig:qram-bb}. Once a readout has been performed the routing network is uncomputed ~\cite{qram}.  \label{fig:bb_circuit}}
\end{figure*}

\begin{figure*}
\begin{center}
    \includegraphics[scale=0.8]{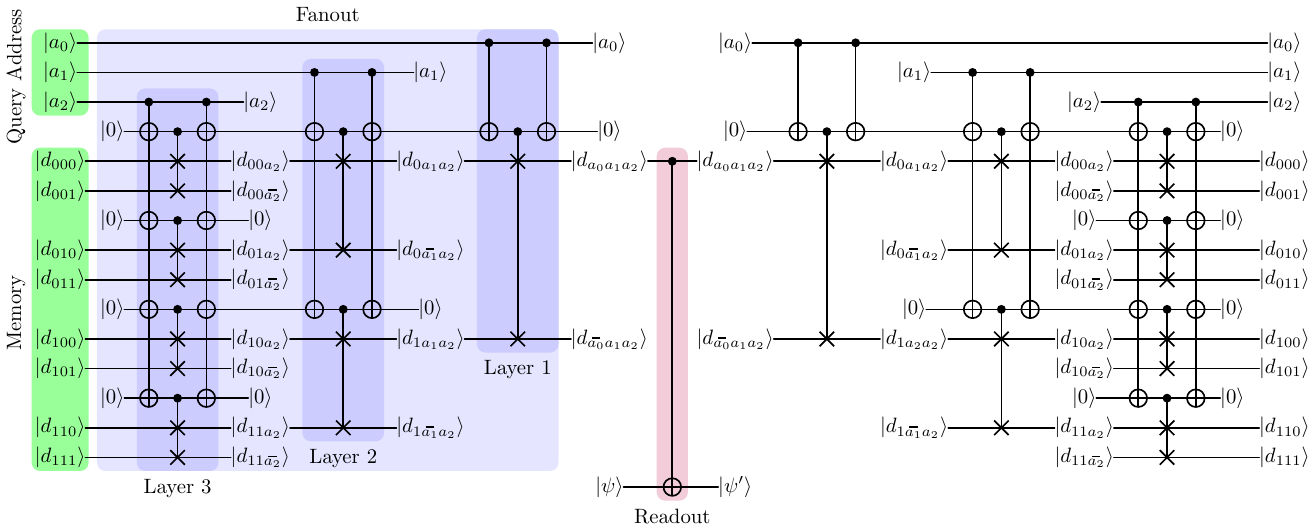}
\end{center}
    \caption{Fanout and SWAP QRAM circuit ~\cite{qram_critique}. \label{fig:fanout_circuit}}
\end{figure*}

\end{document}